\documentclass[prd,aps,showpacs,tightenlines,preprint,nofootinbib,superscriptaddress,floatfix]{revtex4-1}

\usepackage{epsfig}
\usepackage{graphicx}
\usepackage{multirow} 
\usepackage{color}
\usepackage[dvipsnames]{xcolor}
\usepackage{amsmath,amssymb}
\usepackage{rotating}
\usepackage[export]{adjustbox} 
\usepackage[caption=false]{subfig}
\usepackage{floatrow}
\newfloatcommand{capbtabbox}{table}[][\FBwidth]
\usepackage{tabularx}
\usepackage[pdfencoding=auto, psdextra]{hyperref}
\usepackage{units}
\usepackage[T1]{fontenc}
\usepackage[normalem]{ulem} 
\usepackage{aas_macros}

\newcommand{\be}{\begin{equation}}
\newcommand{\ee}{\end{equation}}
\newcommand{\bea}{\begin{eqnarray}}
\newcommand{\eea}{\end{eqnarray}}

\newcommand{\sigmaT}{\sigma_{\mathrm T}}
\newcommand{\sigmaV}{\sigma_{\mathrm V}}

\newcommand\york{Department of Physics and Astronomy, York University,\\Toronto, Ontario M3J 1P3, Canada}
\newcommand\aachen{Institute for Theoretical Particle Physics and Cosmology~(TTK), RWTH Aachen University, 52056 Aachen, Germany}
\newcommand\frankfurt{Institute for Theoretical Physics, Goethe University, 60438 Frankfurt am Main, Germany}

\begin{document}

\title{The Semi-Classical Regime for Dark Matter Self-Interactions}

\author{Brian~Colquhoun}
\email{bcolqu@yorku.ca}
\affiliation{\york}
\author{Saniya Heeba}
\email{heeba@physik.rwth-aachen.de}
\affiliation{\aachen}
\author{Felix~Kahlhoefer}
\email{kahlhoefer@physik.rwth-aachen.de}
\affiliation{\aachen}
\author{Laura~Sagunski}
\email{sagunski@itp.uni-frankfurt.de}
\affiliation{\frankfurt}
\affiliation{\aachen}
\author{Sean~Tulin}
\email{stulin@yorku.ca}
\affiliation{\york}

\date{\today}

\begin{abstract}
Many particle physics models for dark matter self-interactions---motivated to address long-standing challenges to the collisionless cold dark matter paradigm---fall within the semi-classical regime, with interaction potentials that are long-range compared to the de Broglie wavelength for dark matter particles.
In this work, we present a quantum mechanical derivation and new analytic formulas for the semi-classical momentum transfer and viscosity cross sections for self-interactions mediated by a Yukawa potential.
Our results include the leading quantum corrections beyond the classical limit and allow for both distinguishable and identical dark matter particles.
Our formulas supersede the well-known formulas for the momentum transfer cross section obtained from the classical scattering problem, which are often used in phenomenological studies of self-interacting dark matter.
Together with previous approximation formulas for the cross section in the quantum regime, our new results allow for nearly complete analytic coverage of the parameter space for self-interactions with a Yukawa potential.
We also discuss the phenomenological implications of our results and provide a new velocity-averaging procedure for constraining velocity-dependent self-interactions. Our results have been implemented in the newly released code CLASSICS.
\end{abstract}
\preprint{TTK-20-39}
\maketitle

\section{Introduction \label{sec:intro}} 

Collisional interactions among dark matter (DM) particles arise from the underlying microphysics of DM.
If governed solely by weak-scale physics, the scattering rate is sufficiently small that DM behaves as a collisionless fluid during cosmic structure formation.
On the other hand, if the hidden forces between DM particles are comparable to the nuclear forces between protons and neutrons---with self-scattering cross section per unit mass of $\sigma/m \sim {\rm barn \, GeV^{-1}}$---then DM self-interactions affect the inner structure of galactic halos on kiloparsec scales.
In this case, longstanding tensions between observations and $N$-body simulations for cold collisionless DM can be brought into accord through the effect of self-interactions~\cite{Spergel:1999mh} (see Ref.~\cite{Tulin:2017ara} for a review).\footnote{We also refer to Refs.~\cite{Bullock:2017xww,Salucci:2018hqu} for further recent reviews of small scale structure issues for dark matter.}
DM self-interactions may be mediated by new dark sector particles at or below the GeV scale, which are actively being searched for in particle physics experiments worldwide~\cite{Essig:2013lka,Alexander:2016aln,Beacham:2019nyx}.

Self-scattering in DM halos is non-relativistic and the cross section is calculated by solving the Schr\"{o}dinger equation, which provides a non-perturbative resummation of multiple mediator exchanges during a single scattering~\cite{Buckley:2009in,Tulin:2012wi,Tulin:2013teo}.
By analogy with nuclear forces, the most widely considered interaction is the Yukawa potential~\cite{Feng:2009hw,Feng:2010zp,Loeb:2010gj}, expressed in natural units ($\hbar = c = 1$) as
\begin{equation} \label{eq:Yukawa}
U(r)=\pm \frac{\alpha_\chi}{r} e^{-m_\phi r} \, ,
\end{equation}
where $\alpha_\chi$ is the dark sector analog of the fine structure constant. 
The potential $U(r)$ can be attractive ($-$) or repulsive ($+$) depending on the nature of the DM particle(s) $\chi$ and mediator particle $\phi$ (their masses are denoted $m_\chi$ and $m_\phi$, respectively).
The latter can be a fundamental scalar or vector boson in a weakly-interacting theory~\cite{Buckley:2009in,Feng:2009hw,Feng:2010zp,Loeb:2010gj} or a composite state exchanged as a residual force in a strongly-interacting theory~\cite{Boddy:2014yra}.

In general, analytic solutions to the Schr\"{o}dinger equation are not possible for the Yukawa potential outside the Born (perturbative) regime where $\alpha_\chi m_\chi/m_\phi \ll 1$.
However, for momentum $k \ll m_\phi$, $S$-wave scattering dominates and accurate approximations have been obtained for the cross section, both for the Yukawa potential (using the Hulth\'{e}n approximation)~\cite{Tulin:2013teo} and in a general framework~\cite{Chu:2019awd}.
On the other hand, for $k \gtrsim m_\phi$, the partial wave analysis requires a sum over higher modes $\ell > 0$, so the calculation of the cross section becomes numerically intensive for $k \gg m_\phi$, as one approaches the classical limit $k / m_\phi \to \infty$~\cite{Tulin:2012wi,Tulin:2013teo}.

\begin{figure}
  \begin{center}
    \includegraphics[width=0.65\textwidth]{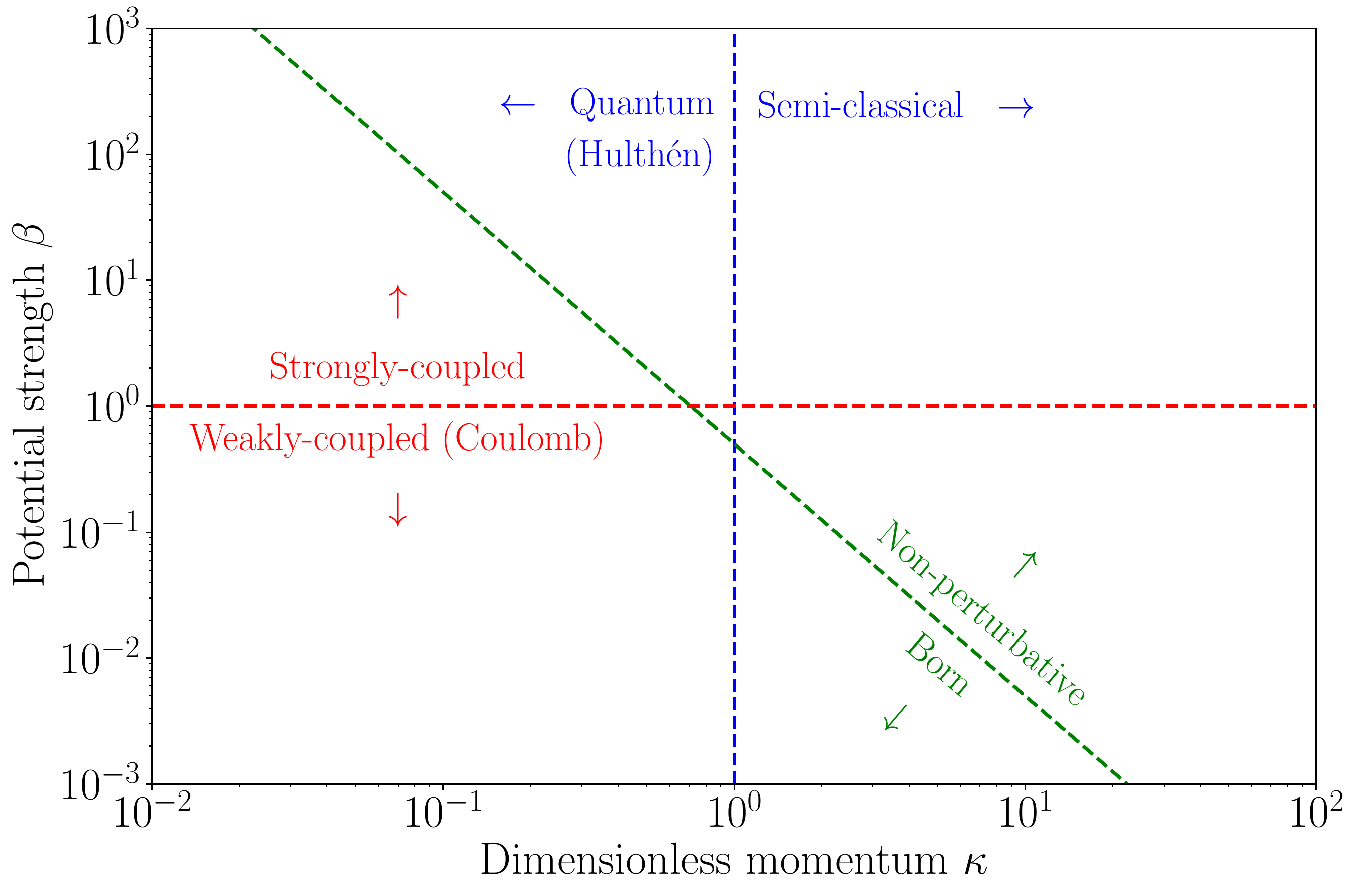}
  \caption{\label{fig:regimes}Sketch of the different regimes relevant for calculating self-interaction cross sections for a Yukawa potential in terms of the dimensionless parameters $\beta$ and $\kappa$. The aim of the present work is to obtain analytical expressions for the scattering cross section in the semi-classical and non-perturbative regime.}
  \end{center}
\end{figure}

The different regimes in this problem are delineated by two dimensionless parameters
\be \label{eq:redefs1}
\kappa = k/m_\phi \, , \quad 
\beta = \frac{2\alpha_\chi m_\phi}{m_\chi v^2} \, ,
\ee
which correspond to the dimensionless momentum and the strength of the potential relative to the kinetic energy, respectively.
In the classical scattering problem, the cross section is parametrized in terms of $\beta$ alone, yielding Coulomb-like scattering for $\beta \ll 1$, while $\beta \gg 1$ corresponds to the case of a strong potential~\cite{PhysRevLett.90.225002,PhysRevE.70.056405}.
We can summarize the different regimes as follows:
\be \label{eq:regimes}
\begin{aligned}
\textnormal{weakly-coupled ($\beta \ll 1$)} \quad &\textnormal{vs.} \quad
\textnormal{strongly-coupled ($\beta \gg 1$)}\,, \\
\textnormal{Born ($2 \beta \kappa^2 \ll 1$)} \quad &\textnormal{vs.} \quad
\textnormal{non-perturbative regime ($2 \beta \kappa^2 \gg 1$)}\,,  \\
\textnormal{quantum ($\kappa \ll 1$)}
\quad &\textnormal{vs.} \quad
\textnormal{semi-classical ($\kappa \gtrsim 1$)}  \; .
\end{aligned}
\ee
Fig.~\ref{fig:regimes} illustrates the landscape of these various cases, each of which involves different approximations for computing the scattering cross section analytically.

Generally, quoting a single figure-of-merit $\sigma/m$ is an over-simplification since the differential cross section $\mathrm{d}\sigma/\mathrm{d}\Omega$ depends nontrivially on both the relative scattering velocity $v$ and scattering angle $\theta$.
The former is treated by phase-space averaging, while for the latter, one uses a suitable proxy for the cross section, as opposed to the total cross section $\sigma = \int \mathrm{d}\Omega (\mathrm{d}\sigma/\mathrm{d}\Omega)$.
In the literature, the most widely-used proxy for $\sigma$ is
the momentum transfer cross section~\cite{Mohapatra:2001sx,Buckley:2009in,Feng:2009hw}
\begin{equation} \label{eq:sigmaT_def}
\sigmaT \equiv \int \mathrm{d}\Omega \, (1-\cos\theta) \,\frac{\mathrm{d}\sigma}{\mathrm{d}\Omega}  \, ,
\end{equation}
where $(1-\cos\theta)$ is the fractional change in longitudinal momentum.
Analytic formulas for $\sigmaT$ for the Yukawa potential in the classical limit have been obtained by solving the classical scattering problem for distinguishable particles, originally motivated in the context of transport theory for plasmas~\cite{PhysRevLett.90.225002,PhysRevE.70.056405} and subsequently imported into a DM context~\cite{Feng:2009hw,Feng:2010zp}.
In the kinetic theory of gases, $\sigmaT$ arises as the relevant quantity for diffusion rates~\cite{1965itpg.book.....V}.
For DM, this quantity governs the effective drag force and evaporation rate felt by a subhalo merging into a static background of DM particles approximating a larger halo~\cite{Kahlhoefer:2013dca}.
However, $\sigmaT$ as defined in Eq.~\eqref{eq:sigmaT_def} is problematic since it preferentially weights backward scattering, which leaves DM distributions unchanged, and moreover is ill-defined for identical particles.\footnote{Ref.~\cite{Kahlhoefer:2013dca} proposed modifying $\sigmaT$ to $\tilde{\sigma}_\mathrm{T} = \int d\Omega \, (1- |\cos \theta|) \frac{d\sigma}{d\Omega}$ to ameliorate these issues. 
Since $|\cos \theta|$ has no simple decomposition in terms of Legendre polynomials, this approach is difficult to incorporate into our quantum treatment to follow and does not lead to simple expressions in terms of phase shifts.}

As an alternative, there is the viscosity cross section~\cite{Tulin:2013teo}
\begin{equation}\label{eq:sigmaV_def}
\sigmaV \equiv \int \mathrm{d} \Omega \, \sin^2 \theta \, \frac{\mathrm{d}\sigma}{\mathrm{d}\Omega} \, ,
\end{equation}
which arises from kinetic theory as the relevant quantity for viscosity and heat conductivity~\cite{1965itpg.book.....V}.
This quantity may be a better figure-of-merit for DM halo dynamics since in fluid formulations of self-interacting DM, the effect of DM scattering is described by heat conduction~\cite{Gnedin:2000ea,Balberg:2002ue}.
Moreover, $\sigmaV$ is well-defined for identical particles and preferentially weights perpendicular scattering, which is the maximal modification of particle trajectories within halos.

The goal of the present work is to provide an analytic and fully quantum mechanical description of the self-interaction cross sections $\sigmaT$ and $\sigmaV$ in the semi-classical regime, which is lacking in the literature.
This has two advantages over the purely classical scattering problem.
First, we are able to retain the leading terms involving factors of $1/\kappa$, which necessarily arise due to the uncertainty principle but are neglected in the classical limit $\kappa \to \infty$.
This allows us to accurately approximate the entire semi-classical regime, $\kappa \gtrsim 1$, which we demonstrate by comparing to exact numerical solutions to the Schr\"{o}dinger equation.
Second, we consider the case of identical particles for $\sigmaV$, which cannot be treated classically.
Our analytic formulas for $\sigmaV$, both for distinguishable and identical particles, are new.

The remainder of this work is organized as follows.
Sec.~\ref{sec:formalism} describes the classical and quantum calculations.
For distinguishable particles, we show that there is a direct correspondence between the two treatments in the limit $\kappa \to \infty$.
We also consider the case of identical particles for which there is no classical analog.
In Sec.~\ref{sec:analytic}, we derive analytic formulas for $\sigmaT$ and $\sigmaV$.
We obtain separate results in the weakly-coupled ($\beta \ll 1)$ and strongly-coupled ($\beta \gg 1$) regimes, which are joined together at $\beta \sim 1$ with an empirical fit function.
We consider cases for both attractive and repulsive Yukawa potentials, as well as both distinguishable and identical particles.
In Sec.~\ref{sec:results}, we present our main results, comparing our analytic formulas with numerical results from solving the Schr\"{o}dinger equation.
We also consider the phenomenological implications for self-interacting DM in the semi-classical regime (and beyond).
Lastly, Sec.~\ref{sec:concl} summarizes our conclusions. Additional details are provided in Appendices~\ref{app:review}--\ref{app:averaging}. 

We make our results available through the new code {\bf C}alcu{\bf LA}tion{\bf S} of {\bf S}elf-{\bf I}nteraction {\bf C}ross {\bf S}ections (CLASSICS).
This code implements analytical expressions for the various cross sections in the semi-classical and quantum regimes---covering the full parameter space for self-interactions with a Yukawa potential---as well as provides tabulated values for velocity-averaged results.
It is available at \url{https://github.com/kahlhoefer/CLASSICS}.

\section{General formalism\label{sec:formalism}} 

\subsection{Classical calculation}

Classically, cross sections are calculated by solving for trajectories to determine the resulting scattering angle $\theta(b)$ for a given impact parameter $b$ (for a review see Ref.~\cite{Heer}).
For a central potential $U(r)$, this is given by the integral
\be \label{eq:scatt_angle}
\theta(b) = \pi - \int_{r_0}^\infty \mathrm{d}r\,   \frac{2b}{r^2\sqrt{1 - U(r)/E - b^2/r^2  }} \; ,
\ee
where $r_0$ is the classical turning point, which is the largest root where the term under the square root vanishes. 
We also have the usual definitions for energy $E = \mu v^2/2$, momentum $k = \mu v$, reduced mass $\mu = m_\chi/2$, and relative velocity $v$.
After solving Eq.~\eqref{eq:scatt_angle}, the differential cross section is expressed as $\mathrm{d}\sigma/\mathrm{d}\Omega = \mathrm{d}^2 b/\mathrm{d}\Omega$ and Eqs.~\eqref{eq:sigmaT_def} and \eqref{eq:sigmaV_def} are 
\begin{subequations}
\label{eq:clas_cross_sections}
\bea 
\sigmaT &=& \int \mathrm{d}\Omega \, (1- \cos\theta) \, \frac{\mathrm{d}^2 b}{\mathrm{d}\Omega} = 2\pi \int_{0}^\infty \mathrm{d}b \, b \left(1-\cos\theta(b)\right)\\
\sigmaV &=& \int \mathrm{d}\Omega \, \sin^2\theta \, \frac{\mathrm{d}^2 b}{\mathrm{d}\Omega} = 2\pi \int_{0}^\infty \mathrm{d}b \, b \sin^2\theta(b) \, .
\eea
\end{subequations}
In the classical calculation, the lower limit of integration in Eqs.~\eqref{eq:clas_cross_sections} is simply $b=0$.
However, as we show below, the corresponding cross sections for quantum mechanical scattering (of non-identical particles) are given by exactly the same formulas in the semi-classical regime, with the simple modification that the lower limit of integration is $b_{\rm min} = \mathcal{O}(\frac{1}{2k}) > 0$.

There is a heuristic argument for $b_{\rm min}$ from the uncertainty principle.
If we assume the initial state has a sufficiently well-defined momentum $k$, then the corresponding uncertainty must satisfy $\Delta k \lesssim k$.
Similarly, the impact parameter $b$ cannot be smaller than the uncertainty in position, $\Delta r$.
By the uncertainty principle, we have
\be
b \gtrsim \Delta r \gtrsim \frac{1}{2\Delta k} \gtrsim \frac{1}{2k} .
\ee
This same result follows from the partial wave analysis below: we find $b_{\rm min} = \frac{1}{2k}$ for $\sigmaT$ and $b_{\rm min} = \frac{1}{k}$ for $\sigmaV$.

For the Yukawa potential~\eqref{eq:Yukawa}, it is useful to re-express the previous equations in terms of dimensionless parameters
\be \label{eq:redefs}
R = r m_\phi \, , \quad
\rho = b m_\phi 
\, .
\ee
Then, including the appropriate lower limits $b_{\rm min}$, Eqs.~\eqref{eq:clas_cross_sections} become
\be \label{eq:clas_cross_sections_2}
\sigma_\mathrm{T} m_\phi^2 =  2\pi \int_{1/2\kappa}^\infty \mathrm{d}\rho \, \rho \left(1-\cos\theta(\rho)\right) \, ,
\qquad 
\sigma_\mathrm{V} m_\phi^2 = 2\pi \int_{1/\kappa}^\infty \mathrm{d}\rho \, \rho \sin^2\theta(\rho) \, ,
\ee
where
\be \label{eq:scat_angle}
\theta(\rho) = \pi - \int_{R_0}^\infty \mathrm{d}R \,  \frac{2\rho}{R^2 \sqrt{1 \mp 2\beta e^{-R}/R  - \rho^2/R^2  }}
\ee
with the upper (lower) sign for a repulsive (attractive) potential and the classical turning point $R_0 = r_0 m_\phi$.
The classical calculation corresponds to the limit $\kappa \to \infty$, in which the lower limits of Eqs.~\eqref{eq:clas_cross_sections_2} are replaced by zero, yielding cross sections that are a function of a single variable $\beta$~\cite{PhysRevLett.90.225002,PhysRevE.70.056405}. This approach has been pursued in detail in the plasma physics literature and we briefly summarize the central results in Appendix~\ref{app:review}.
However, these results are only valid in the limit $\kappa \to \infty$, while more generally in the semi-classical regime, the cross sections now depend on two parameters, $\beta$ and $\kappa$.

\subsection{Quantum calculation}

In the usual partial wave analysis, the asymptotic form of the radial wavefunction $\mathcal{R}_\ell(r)$ is determined by solving the Schr\"{o}dinger equation
\be \label{eq:Schr_1}
\frac{1}{r^2} \frac{\partial}{\partial r} \left( r^2 \frac{\partial \mathcal{R}_\ell}{\partial r} \right) + \left( E - U(r) - \frac{\ell(\ell+1)}{2\mu r^2}  \right) \mathcal{R}_\ell(r) = 0 \, .
\ee
Rescaling the wavefunction as $u_\ell(R) = r \mathcal{R}_\ell(r)$, with the definitions in Eq.~\eqref{eq:redefs}, the Schr\"{o}dinger equation becomes
\be
\frac{d^2 u_\ell}{dR^2} + K^2(R)\,  u_\ell(R) = 0 \, ,
\ee
where 
\be
K^2(R) = \kappa^2 \left( 1 - \frac{\ell(\ell+1)}{\kappa^2 R^2} \mp \frac{2 \beta}{R} e^{-R} \right)\,.
\ee
The phase shifts $\delta_\ell$ are obtained from the asymptotic solutions for the wavefunction, 
\be \label{eq:uR_largeR}
u_\ell(R) \xrightarrow[R \to \infty]{} \sin\left(\kappa R - \frac{\ell\pi}{2} + \delta_\ell\right) \, ,
\ee
up to an overall (irrelevant) normalization.
Given the phase shifts, the differential cross section is given by the usual formula
\begin{equation} 
 \frac{\mathrm{d}\sigma}{\mathrm{d}\Omega} = \frac{1}{k^2} \left| \sum_{\ell=0}^\infty (2\ell + 1) e^{i\delta_\ell} P_{\ell}(\cos \theta) \sin \delta_\ell \right|^2 \; .
\end{equation}
By integrating over angle, the transfer and viscosity cross sections can be expressed as a sum over partial wave modes
\begin{subequations} \label{eq:sigma_TV_sum}
\bea
\sigmaT m_\phi^2 &=& \frac{4\pi}{\kappa^2} \sum_{\ell=0}^\infty (\ell+1) \sin^2 (\delta_{\ell+1} - \delta_{\ell}) \, , 
\label{eq:sigma_T_sum} \\
\sigmaV m_\phi^2 &=& \frac{4\pi}{\kappa^2} \sum_{\ell=0}^\infty \frac{(\ell+1)(\ell+2)}{2\ell+3}\sin^2 (\delta_{\ell+2} - \delta_{\ell}) \; . 
\eea
\end{subequations}
From the preceding formulas, it is clear that the quantum calculations of Eqs.~\eqref{eq:sigma_TV_sum} depend in general on the two independent parameters, $\kappa$ and $\beta$, while the classical calculation only depends on $\beta$.

In the semi-classical approximation~\cite{Berry:1972na}, the wavefunction is approximated using the WKB form
\be \label{eq:WKB}
u_\ell^{\rm WKB}(R) = \frac{2}{K(R)^{1/2}} \cos\left( \int_{R_0}^R dR^\prime K(R^\prime) - \frac{\pi}{4} \right) \, .
\ee
The phase shift $\delta_\ell$ is simply the difference between the phase of Eq.~\eqref{eq:WKB} and the phase when $\beta=0$, as $R \to \infty$:
\be
\delta_\ell = \lim_{R\to \infty} \left( \int_{R_0}^R dR^\prime K(R^\prime)- \frac{\pi}{4} \right) - \left( \int_{R_0}^R dR^\prime K(R^\prime) - \frac{\pi}{4} \right)_{\beta=0} \, .
\ee
However, there is a well-known problem with Eq.~\eqref{eq:WKB} as is, which is made manifest by considering $\beta=0$.
In this case, the free WKB wavefunction is asymptotically (up to an overall constant)
\be \label{eq:WKB_largeR}
u_\ell^{\rm WKB}(R) \xrightarrow[R \to \infty]{} \sin\left(\kappa R - \frac{\pi}{2}\sqrt{\ell(\ell+1)} + \frac{\pi}{4} \right) \, .
\ee
This does not agree with the expectation for the free wavefunction, $u_\ell(R) \to \sin(\kappa R - \tfrac{\ell\pi}{2})$, unless one makes the approximation $\ell \gg 1$.

This issue is solved by taking an alternative form for the centrifugal potential, replacing $\ell(\ell+1) \to (\ell + \tfrac{1}{2})^2$.
The WKB wavefunction is thereby rescued provided one substitutes
\be \label{eq:Langer_sub}
K^2(R) \longrightarrow K_{\rm eff}^2(R) = \kappa^2 \left( 1 - \frac{(\ell+\tfrac{1}{2})^2}{\kappa^2 R^2} \mp \frac{2\beta}{R} e^{-R} \right) 
\ee
in Eq.~\eqref{eq:WKB}, which yields the correct asymptotic behavior for any $\ell$.
Originally discovered by Kramers~\cite{Kramers:1926njj} and others as a fix to reproduce various known exact results (see Ref.~\cite{Berry:1972na}), it was put on firm theoretical footing by Langer~\cite{Langer:1937qr} and is known in the literature as the Langer modification. 
The argument is that the WKB condition breaks down where $K^2(R)$ has singular points, namely at $R=0$ due to the centrifugal term.
Remarkably, however, through a change of variables, one can transform the problem into one for which the WKB condition is satisfied over the entire domain; upon transforming back to the original variables, this simply amounts to the substitution $\ell(\ell+1) \to (\ell + \tfrac{1}{2})^2$~\cite{Langer:1937qr}.
The semi-classical phase shift, after substituting Eq.~\eqref{eq:Langer_sub}, is now
\be
\delta_\ell = \kappa \int^\infty_{R_0} dR \, \left( \sqrt{ 1 \mp \frac{2\beta}{R} e^{-R} - \frac{(\ell+\tfrac{1}{2})^2}{\kappa^2 R^2} } - 1 \right) + \frac{\pi(\ell+\tfrac{1}{2})}{2} - \kappa R_0 \, ,
\label{eq:deltalexact}
\ee
where $R_0$ is also determined from $K_{\rm eff}^2$.

Next, we turn to the cross sections in Eqs.~\eqref{eq:sigma_TV_sum}.
Since a large number of $\ell$ modes are summed over, it is possible to interpret $\delta_\ell$ as a continuous function $\delta(\ell)$ and approximate
\begin{subequations}
\bea
\label{eq:deltaprime} \delta_{\ell+1} - \delta_{\ell} & \approx \delta'(\ell + 1/2) \, , \\
  \delta_{\ell+2} - \delta_{\ell} & \approx 2 \delta'(\ell + 1) \; ,
\eea
  \end{subequations}
where 
\be \label{eq:delta_derivative}
\delta'(\ell) = 
\frac{\pi}{2} - \int_{R_0}^\infty dR \,  \frac{\ell+\tfrac{1}{2}}{\kappa R^2 \sqrt{1 \mp \frac{2\beta}{R} e^{-R}  - \frac{(\ell+1/2)^2}{\kappa^2 R^2}  } } \; .
\ee
Using this approximation, we can
turn the discrete sums into integrals. While there is no unique way to do this, we find that numerically obtained results (see below) can be accurately reproduced when defining the integrals in such a way that the discrete sums are recovered when approximating the integrals by a Riemann sum with partition size $\Delta \ell = 1$ using the midpoint rule:\footnote{According to this approximation, an integral of the form $I = \int_{\ell_\text{min}}^\infty f(\ell) \mathrm{d}\ell$ can be written as
\begin{equation}
 I \approx \sum_{i=0}^\infty f\left(\ell_\text{min} + \left(i + \tfrac{1}{2}\right)\Delta \ell\right) \Delta \ell =  \sum_{\ell=0}^\infty f\left(\ell_\text{min} + \ell + \tfrac{1}{2}\right) \; , \nonumber
\end{equation}
where in the second step we have taken $\Delta\ell = 1$. Taking, for example, $\ell_\text{min} = 0$ and $f(\ell) = (\ell + 1/2) \sin^2 \delta'(\ell)$ as in Eq.~\eqref{eq:sigmaTinteg} then yields Eq.~\eqref{eq:sigma_T_sum} with the approximation from Eq.~\eqref{eq:deltaprime}. While the midpoint rule yields a sufficient level of accuracy for the present study, more sophisticated treatments were discussed in Ref.~\cite{Berry:1972na}.
}
\begin{subequations} \label{eq:sigma_TV_int}
\bea
\sigmaT m_\phi^2 &\approx& \frac{4\pi}{\kappa^2} \int_{0}^\infty \mathrm{d}\ell \, (\ell + \tfrac{1}{2}) \, \sin^2 \delta'(\ell) \nonumber \\
& = & \frac{4\pi}{\kappa^2} \int_{1/2}^\infty \mathrm{d}\ell \, \ell \, \sin^2 \delta'(\ell - 1/2)  \, , \label{eq:sigmaTinteg}\\
\sigmaV m_\phi^2 &\approx& \frac{4\pi}{\kappa^2} \int_{0}^\infty \mathrm{d}\ell \, \frac{(\ell + 1/2)(\ell + 3/2)}{2\ell + 2} \, \sin^2  2\delta'(\ell + 1/2) \nonumber \\
&\approx& \frac{2\pi}{\kappa^2} \int_{1}^\infty \mathrm{d}\ell \, \ell \, \sin^2 2\delta'(\ell - 1/2) \; \label{eq:sigmaVinteg},
\eea
\end{subequations}
where we have used the approximation $(\ell+1/2)(\ell+3/2)/(2\ell+2) \approx (\ell + 1)/2$ in the final step. Equating $\ell + \tfrac{1}{2} = \kappa \rho$ and comparing Eqs.~\eqref{eq:scat_angle} and \eqref{eq:delta_derivative}, it is clear that $\delta'(\ell)$ is simply $\theta(\rho)/2$ and therefore Eqs.~\eqref{eq:sigma_TV_int} are identical to Eqs.~\eqref{eq:clas_cross_sections} where $b_{\rm min} = \tfrac{1}{2\kappa}$ ($\tfrac{1}{\kappa}$) for $\sigmaT$ ($\sigmaV$).
This comes from the extra term from the Langer modification. 

\subsection{Identical particles}

A crucial advantage of the quantum mechanical treatment is that we can directly extend our discussion to the scattering of identical particles. For identical bosons, the total wave function needs to be symmetric under particle exchange, so the spatial part needs to be even (odd) if the spin part is symmetric (antisymmetric). For identical fermions, a symmetric (antisymmetric) spin part implies an odd (even) spatial part of the wave function. The differential cross section is then given by
\begin{equation}
\frac{\mathrm{d} \sigma}{\mathrm{d}\Omega} = \frac{1}{k^2} \left\lvert \sum_{\ell=0}^\infty(2\ell+1) e^{i \delta_\ell} \left[P_\ell(\cos \theta) \pm P_\ell(-\cos \theta) \right] \sin \delta_\ell \right\rvert^2 \; ,
\end{equation}
where the positive sign corresponds to an even spatial wave function and the negative sign to an odd one. Since $P_\ell(-x) = (-1)^\ell P_\ell(x)$ it follows that the contribution from all odd phase shifts vanishes for the even case, while the contribution from even phase shifts vanishes for the odd case.

Hence we obtain in both cases
\begin{equation}
 \frac{\mathrm{d} \sigma}{\mathrm{d}\Omega}(\theta) = \frac{\mathrm{d} \sigma}{\mathrm{d}\Omega}(\pi - \theta) \; .
\end{equation}
It follows immediately that $\int \mathrm{d} \cos \theta  \tfrac{\mathrm{d}\sigma}{\mathrm{d}\Omega} \cos \theta = 0$ and hence $\sigma_\mathrm{T} = \sigma$. In other words, the momentum transfer cross section is no longer useful for the scattering of identical particles.
The most relevant quantity for the scattering of identical particles is hence the viscosity cross section, which is given by
\begin{align}
 \sigma_\mathrm{V}^\text{even} m_\phi^2 & = \frac{8\pi}{\kappa^2} \sum_{\ell = 0}^\infty \frac{(2\ell + 1)(2\ell + 2)}{4\ell+3}\sin^2(\delta_{2\ell+2} - \delta_{2\ell}) \, , \label{eq:bosonic}\\
 \sigma_\mathrm{V}^\text{odd} m_\phi^2 & = \frac{8\pi}{\kappa^2} \sum_{\ell = 0}^\infty \frac{(2\ell+2)(2\ell + 3)}{4\ell+5}\sin^2(\delta_{2\ell+3} - \delta_{2\ell+1}) \; ,
\end{align}
where an overall symmetry factor $\tfrac{1}{2}$ has been introduced to avoid double-counting. 
Note that it follows directly that
\begin{equation}
  \sigma_\mathrm{V}^\text{even} +  \sigma_\mathrm{V}^\text{odd} = 2 \sigma_\mathrm{V} \label{eq:boson-fermion}\; .
\end{equation}
If a large number of phase shifts give a relevant contribution to the viscosity cross section, one finds
\begin{equation}
 \sigma_\mathrm{V}^\text{even} \approx  \sigma_\mathrm{V}^\text{odd}  \approx \sigma_\mathrm{V} \; .
\end{equation}
If, on the other hand, the cross section is dominated by the first few phase shifts, we expect $ \sigma_\mathrm{V}^\text{even} \gg \sigma_\mathrm{V}^\text{odd}$.

Again we can approximate the cross sections for the scattering of identical particles through integrals, making use of the same prescription as above:
\begin{align}
 \sigma_\mathrm{V}^\text{even} \, m_\phi^2 & \approx \frac{8\pi}{\kappa^2} \int_{0}^\infty \mathrm{d}\ell\, \frac{(2\ell)(2\ell+1)}{4\ell+1} \sin^2 2\delta'(2\ell)\nonumber \\
 & \approx \frac{2\pi}{\kappa^2} \int_{1/2}^\infty\mathrm{d}\ell \, \ell \sin^2 2\delta'(\ell-1/2)\, ,\\ 
 \sigma_\mathrm{V}^\text{odd} \, m_\phi^2 & \approx \frac{8\pi}{\kappa^2} \int_{0}^\infty \mathrm{d}\ell \, \frac{(2\ell+1)(2\ell+2)}{4\ell+3} \sin^2 2\delta'(2\ell+1) \nonumber \\
 & \approx \frac{2\pi}{\kappa^2} \int_{3/2}^\infty  \mathrm{d}\ell \, \ell \sin^2 2\delta'(\ell-1/2)\; .
\end{align}
As we will see below, in certain situations the argument of the sine is sufficiently small to allow for the approximation $\sin^2(x) \approx x^2$. Whenever this is the case, it follows that $\sigma_\mathrm{V}^\text{even} \approx 2 \sigma_\mathrm{T}$, i.e.\ the viscosity cross section for the even case is directly related to the momentum transfer cross section for non-identical particles.

Finally, we note that for DM particles with non-zero spin the more realistic situation is that the scattering particles are unpolarized and hence the total spin wave function may be either symmetric or anti-symmetric. Averaging over spins then amounts to the following replacement
\be
\sigmaV \to \begin{cases}
\sigmaV^{\rm even} & \text{scalar DM} \, , \\
\frac{1}{4} \sigmaV^{\rm even} + \frac{3}{4} \sigmaV^{\rm odd} & \text{fermion DM} \; , \\
\frac{2}{3} \sigmaV^{\rm even} + \frac{1}{3} \sigmaV^{\rm odd} & \text{vector DM} \; . \end{cases}
\ee

\section{Analytical approximations for the semi-classical regime}
\label{sec:analytic}

In this section, we derive analytical approximations for the semi-classical regime in a weakly- and strongly-coupled theory, corresponding to  $\beta \ll 1$ and  $\beta \gg 1$, respectively.

\subsection{Weak potential}

Let us first consider a weak dimensionless potential with $\beta \ll 1$. We begin by considering the case that not only is $\beta$ small but also $\kappa \beta \ll 1$. In this case, we can expand the square root in Eq.~\eqref{eq:deltalexact} in $\beta$ and approximate $R_0 \approx (\ell+\tfrac{1}{2})/\kappa$. The phase shift is then given by
\begin{align}
 \delta_\ell \approx \mp \int_{R_0}^\infty \mathrm{d}{R} \frac{\beta \kappa}{R}e^{-R} \frac{1}{\sqrt{1 - \frac{(\ell+\tfrac{1}{2})^2}{\kappa^2 R^2}}} = \mp \beta \kappa K_0\left(\frac{\ell + \tfrac{1}{2}}{\kappa}\right)\,,
 \label{eq:phase-shift_smallbeta} 
\end{align}
where $K_i$ denote the modified Bessel functions of the second kind and the negative (positive) sign corresponds to an attractive (repulsive) potential. For the difference between two consecutive phase shifts, we then obtain
\begin{equation}
 \delta_{\ell} - \delta_{\ell-1} \approx \delta'(\ell-1/2) \approx \pm \beta K_1\left(\frac{\ell}{\kappa}\right)\,.
\end{equation}
Substituting this expression into Eq.~\eqref{eq:sigmaTinteg}, we obtain for both attractive and repulsive potentials
\begin{equation}
 \sigma_\mathrm{T} \approx \frac{4 \pi}{\kappa^2 m_\phi^2} \int_{1/2}^\infty \ell \sin^2\left[ \beta K_1 \left( \frac{\ell}{\kappa}\right) \right] \mathrm{d}\ell\;.
\label{eq:sigma_T}
\end{equation}

For $\beta \kappa < 1/2$ and $\ell > 1/2$ we can make use of the fact that $K_1(x)$ is a monotonically decreasing function to show that
\begin{equation}
 \beta K_1 \left( \frac{\ell}{\kappa}\right) < \beta K_1 \left( \frac{1}{2\kappa}\right) < \beta K_1(\beta) < 1 \; .
\end{equation}
Hence the argument of the sine is small across the entire integration range and we obtain approximately
\begin{align}
\sigma_\mathrm{T}  & \approx \frac{\pi \beta^2}{2\kappa^2 m_\phi^2} \left[ -K_1\left(\frac{1}{2\kappa}\right)^2 + K_0\left(\frac{1}{2\kappa}\right) K_2\left(\frac{1}{2\kappa}\right)\right] \equiv \frac{2\pi \beta^2}{m_\phi^2} \eta\left(\frac{1}{2 \kappa}\right)\; ,
\label{eq:sigmaTsmallBeta1}
\end{align}
where the second step defines the function $\eta(x)$. For $\kappa \gg 1$ this expression can be approximated as
\begin{equation}
\eta\left(\frac{1}{2\kappa}\right) \approx 2 \log 4 \kappa - 1 - 2 \gamma_\mathrm{E} + \frac{1-\gamma_\mathrm{E}+\log 4 \kappa}{4\kappa^2} \; ,
\end{equation}
where $\gamma_\mathrm{E}$ is Euler's constant. We emphasize that the explicit dependence on $\kappa$ is a direct consequence of the fact that the integration range in Eq.~\eqref{eq:sigma_T} does not extend to $\ell = 0$.

Starting from Eq.~\eqref{eq:sigmaVinteg}, a completely analogous calculation for the viscosity cross section yields
\begin{align}
 \sigma_\mathrm{V} & \approx \frac{2 \pi}{\kappa^2 m_\phi^2} \int_{1}^\infty \ell \sin^2\left[ 2\beta K_1 \left( \frac{\ell}{\kappa}\right) \right] \mathrm{d}\ell \approx \frac{4\pi\beta^2}{m_\phi^2} \eta\left(\frac{1}{\kappa}\right)  \; . 
\end{align}

Let us now consider the case that $\beta \ll 1$ but $\kappa \beta > 1/2$. In this case the expansion in Eq.~\eqref{eq:phase-shift_smallbeta} is only a good approximation for $\ell > \ell_\text{min} = \kappa \beta$. Indeed, for small $\ell$ the argument of the sine function becomes large and hence the integrand rapidly oscillates between 0 and 1. We can make a rough estimate of the resulting contribution to the total cross section by approximating $\sin^2(x) \approx 1/2$ for $\ell < \ell_\text{min}$, which gives
\begin{equation}
 \frac{4\pi}{\kappa^2 m_\phi^2} \int_{1/2}^{\ell_{\text{min}}} \ell \sin^2\left( \delta_\ell - \delta_{\ell-1} \right) \mathrm{d}\ell \approx  \frac{\pi}{\kappa^2 m_\phi^2} \left(\ell_{\text{min}}^2 - \frac{1}{4}\right)\;.
\end{equation}
For $\ell > \ell_\text{min}$ we can use the same approximations as before to obtain
\begin{align}
 \frac{4\pi}{\kappa^2 m_\phi^2} \int_{\ell_\text{min}}^\infty \ell \sin^2\left[ \beta K_1\left(\frac{\ell}{\kappa}\right) \right] \mathrm{d}\ell 
 \approx \frac{2\pi \beta^2}{m_\phi^2} \eta\left(\beta\right)\; ,
\end{align}
and hence
\begin{align}
 \sigma_\mathrm{T} \approx \frac{2\pi\beta^2}{m_\phi^2} \left[ \frac{\ell_\text{min}^2 - \tfrac{1}{4}}{2 \kappa^2 \beta^2} + \eta\left(\beta\right)\right]   \; .
\end{align}
If $\beta \ll 1$ while $\kappa \beta \gg 1$, the momentum transfer cross section becomes a function of $\beta$ only:
\begin{align}
 \sigma_\mathrm{T} & \approx \frac{2\pi\beta^2}{m_\phi^2} \left[2 \log \frac{2}{\beta} - \frac{1}{2} - 2 \gamma_\mathrm{E} + \beta^2 \left(1-\gamma_\mathrm{E}+\log \frac{2}{\beta} \right)\right]\; . 
 \label{eq:testing}
\end{align}
When $\beta$ is small enough that we can approximate the square bracket by $2\log \beta^{-1}$, this result agrees with the one obtained in the classical approach. However, for constant $\kappa$ the assumption $\kappa \beta > 1$ necessarily breaks down as $\beta \to 0$ and $\sigma_\mathrm{T}$ starts to decrease more steeply, i.e.\ it scales proportional to $\beta^2$ rather than proportional to $\beta^2 \log \beta^{-1}$. This observation agrees with what is found by solving the Schr\"{o}dinger equation numerically.

We note that the cases $\kappa \beta < 1/2$ and $\kappa \beta > 1/2$ can be combined by defining 
\begin{equation}
 \zeta_n(\kappa,\beta) \equiv \frac{\text{max}(n,\beta\kappa)^2 - n^2}{2 \kappa^2 \beta^2} + \eta\left(\frac{\text{max}(n,\beta\kappa)}{\kappa}\right) \; ,
\end{equation}
which leads to
\begin{align}
\sigma_\mathrm{T} \approx  \frac{2\pi\beta^2}{m_\phi^2} \zeta_{1/2}\left(\kappa,\beta\right)\; .
\end{align}

For the viscosity cross section the integration starts at $\ell = 1$ and we require $2 \beta K_1(\ell/\kappa) < 1$ in order for the argument of the sine to be small. Hence we obtain
\begin{align}
\sigma_\mathrm{V} \approx \frac{4\pi\beta^2}{m_\phi^2} \zeta_1\left(\kappa,2\beta\right) \; .
\end{align}
Finally, for the case of identical particles, we find
\begin{align}
    \sigma_{V}^\text{even} & \approx \frac{4\pi\beta^2}{m_\phi^2} \zeta_{1/2}\left(\kappa,2\beta\right) \; , \\
    \sigma_{V}^\text{odd} & \approx \frac{4\pi\beta^2}{m_\phi^2} \zeta_{3/2}\left(\kappa,2\beta\right) \; .
\end{align}
For $\kappa \beta \ll 1$ (and $n$ of order unity) one finds $\zeta_n\left(\kappa,\beta\right) \approx \eta(n / \kappa)$ independent of $\beta$ and hence 
$\sigma_\mathrm{V}^\text{even} \approx 2 \sigma_\mathrm{T} \gg \sigma_\mathrm{V}^\text{odd}$, while for $\kappa \beta \gg 1$ one finds $\zeta_n\left(\kappa,\beta\right) \approx \tfrac{1}{2} + \eta(\beta)$ independent of $\kappa$ and $n$ and therefore 
$\sigma_\mathrm{V}^\text{even} \approx \sigma_\mathrm{V}^\text{odd} \approx \sigma_\mathrm{V}$.

\subsection{Strong potential}

Let us now turn to the case of large $\beta$, corresponding to a strong dimensionless potential. 
We define an effective potential
\begin{equation}
\label{eq:ueff}
U_\mathrm{eff}(R) \equiv  \left(\frac{\ell + 1/2}{\kappa}\right)^2\frac{1}{R^2} \mp 2\beta\frac{e^{-R}}{R} 
\end{equation}
where the upper (lower) sign is for an attractive (repulsive) interaction.
The largest root of $U_\mathrm{eff}(R)=1$ determines the distance of closest approach $R_0$. 
For large $\beta$, this quantity is decisive for the calculation of the phase shifts and therefore the cross sections, as we show below. 
Since $R_0$ depends on the sign of the potential, we obtain different results for attractive and repulsive cases and therefore need to discuss them separately.
This is in contrast to the weak potential regime, where $R_0 \approx (\ell + \tfrac{1}{2})/\kappa$ and our results were independent of sign.

The calculations in the remainder of this subsection proceed as follows. 
For each potential (attractive or repulsive), we determine an analytic form for the phase shift derivative 
by estimating $R_0$ and
specifying the range of $\ell$ that contributes the most to the cross-section. 
The momentum-transfer and viscosity cross sections are then simply the integrals of these phase shifts over the $\ell$ range of interest, which is given by $0 < \ell < \ell_{\rm max}$ where the upper limit $\ell_{\rm max}$ is defined below. (For the repulsive viscosity cross section, it is necessary to integrate beyond $\ell_{\rm max}$, as we discuss below.)

The lower bound of integration $\ell=0$ is taken from the following argument. 
Numerically evaluating the integrals in Eqs.~\eqref{eq:sigmaTinteg} and~\eqref{eq:sigmaVinteg}, 
we find that the dependence on $\kappa$ is much weaker for large $\beta$ than for small $\beta$.  
Therefore, we will neglect this dependence and assume $\kappa \gg 1$, which implies $\ell \gg 1$. 
Since we can approximate $\ell - 1/2 \approx \ell +1/2 \approx \ell$, we take the lower limit of integration to be $\ell = 0$. 

\subsubsection{Attractive potential}
\label{sec:largebeta_att}

\begin{figure}
  \begin{center}
    \includegraphics[width=0.9\textwidth]{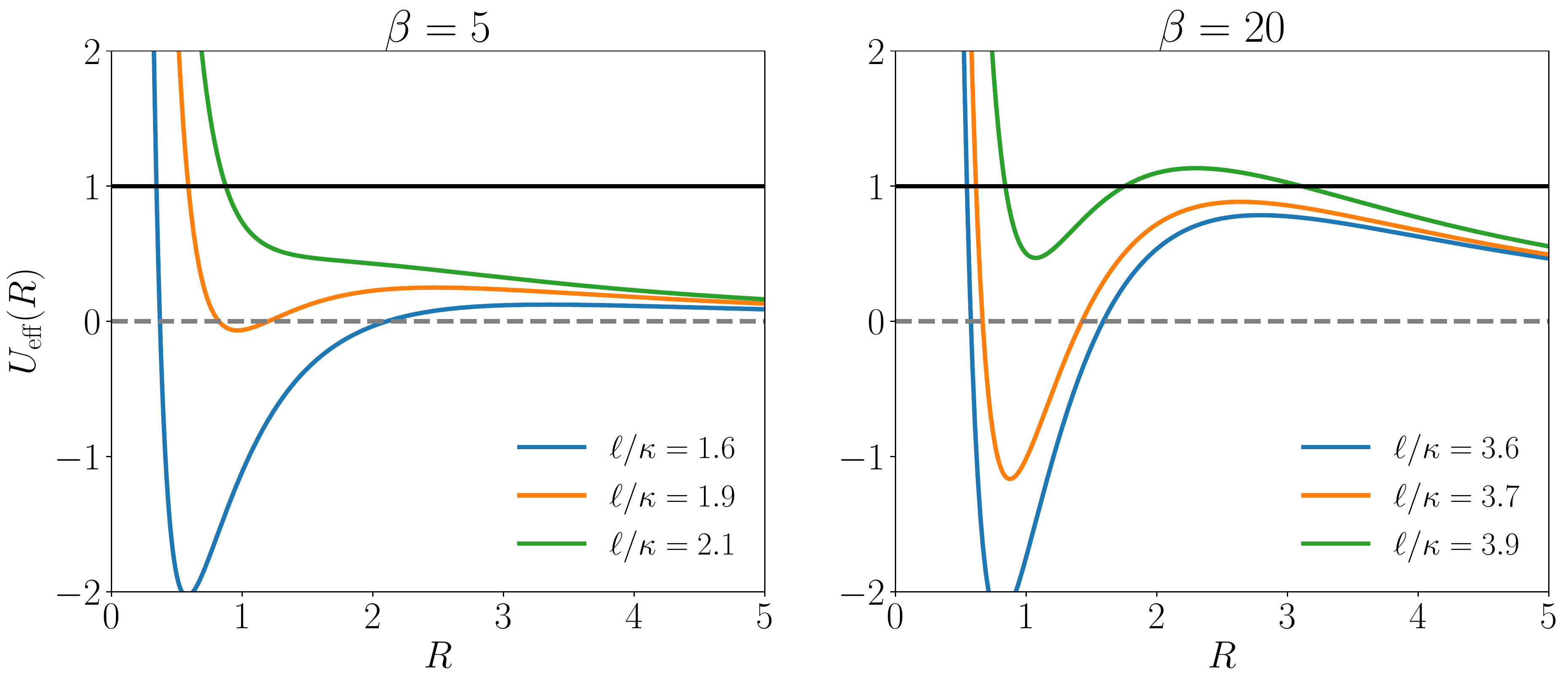}
  \caption{\label{fig:Ueff} Effective potential as described in Eq.~\eqref{eq:ueff} for two values of $\beta$. For $\beta \gtrsim 13.2$, a secondary potential barrier emerges for sufficiently large values of $\ell$, corresponding to a jump in the distance of closest approach. See text for details.}
  \end{center}
\end{figure}

In the attractive case the effective potential $U_\text{eff}(R)$ is monotonically decreasing only for a sufficiently large angular momentum barrier (i.e., large $\ell$). For small $\ell$, on the other hand, the effective potential develops a maximum at finite $R$ which can lead to a secondary potential barrier (see Fig. \ref{fig:Ueff}). For $\beta < 13.2$ this effect is sufficiently small that the equation $U_\text{eff}(R) = 1$ still has a unique solution for all values of $\ell$. For larger $\beta$, however, there will be multiple roots for sufficiently large values of $\ell$. For a given $\beta$ we define $\ell_\text{max}$ as the largest value of $\ell$ for which there is a single root. We show in Appendix~\ref{app:Rmin} that
\begin{align}
\ell_\text{max} \approx \kappa\,\left(1 +\log\beta - \frac{1}{2\,\log\beta}\right)\; .
\end{align}
Further, we find that for $\ell > \ell_{\text{max}}$, the distance of closest approach, $R_0$, is so large that the resulting  phase shifts are tiny and give a negligible contribution to the cross section (see Fig.~\ref{fig:Ueff}). We can therefore approximate
\begin{equation}
    \sigma_\mathrm{T}^\text{att.} m_\phi^2 \approx \frac{4\pi}{\kappa^2} \int_0^{\ell_\text{max}} \ell \sin^2\delta'(\ell) \mathrm{d}\ell \; .
\end{equation}
In order to calculate the derivative of the phase shift, we first estimate $R_0$.
For $\ell < \ell_\text{max}$, one finds $\ell / \kappa < 2\beta$ and hence the equation $U_\text{eff}(R) = 1$ can only be satisfied for $R < 1$. In this case we can approximate $e^{-R} \approx 1$, i.e., the Yukawa potential can be approximated by a Coulomb potential. We then find
\begin{equation}
R_0 \approx \frac{\ell^2}{2\beta\kappa^2} \qquad (\ell < \ell_{\text{max}})\,.
\end{equation}

For completeness, we also calculate $R_0$ in the opposite case ($\ell > \ell_\text{max}$). Here, the (largest) root corresponds to the case that the exponential suppression of the Yukawa potential becomes relevant, which implies $R \gg 1$, as expected. In this case we can ignore the term proportional to $\beta$ in Eq.~\eqref{eq:ueff} to get 
\begin{equation}
R_0 \approx \frac{\ell}{\kappa}  \qquad (\ell > \ell_{\text{max}})\,.
\end{equation}
With these expressions for $R_0$, we can now calculate the phase shifts and their differences.

For $\ell \ll \ell_{\text{max}}$, we show in Appendix~\ref{app:phase_shifts} that the phase shift derivative can be written as
\begin{equation}
    \delta'(\ell) \approx -\frac{\pi}{2} - \frac{\ell}{\kappa} \gamma(\beta) \; ,
    \label{eq:ddeltal_att}
\end{equation}
where
\begin{equation}
    \gamma(\beta) \equiv \frac{W(2 \beta) + 2}{W(2 \beta)^2}
\end{equation}
with $W(x)$ denoting the product logarithm. 
In practice, we find that Eq.~\eqref{eq:ddeltal_att} gives a good approximation for $\tfrac{\ell}{\kappa} \gamma(\beta) < \pi/2$. By defining the rescaled angular momentum $\ell' \equiv \tfrac{\ell}{\kappa} \gamma(\beta)$, we can then write
\begin{equation}
    \sigma_\mathrm{T}^\text{att.} m_\phi^2 \approx \frac{4\pi}{\gamma(\beta)^2} \int_0^{\pi/2} \ell' \sin(\pi/2 + \ell')^2 \mathrm{d}\ell' 
    + 
    \frac{4\pi}{\kappa^2} \int_{\pi \kappa / (2 \gamma(\beta))}^{\ell_{\text{max}}} \ell \sin^2 \delta'(\ell) \mathrm{d}\ell 
     \; .
\end{equation}
The first integral gives $(\pi^2 - 4)/16$. In the second integral there is no simple expression for the derivative of the phase shifts, but since the argument of the sine is large we can approximately replace $\sin^2(x) \approx 1/2$, giving $\ell_\text{max}^2 / 4 - \pi^2 \kappa^2/ (16 \gamma(\beta)^2)$. Hence,
\begin{equation}
    \sigma_\mathrm{T}^\text{att.} m_\phi^2 \approx \frac{\pi}{\kappa^2} \left(\ell_\text{max}^2 - \frac{\kappa^2}{\gamma(\beta)^2}\right) \approx 
    \pi \left[\left(1 +\log\beta - \frac{1}{2\,\log\beta}\right)^2 - \frac{1}{\gamma(\beta)^2}\right]
    \;.
\end{equation}

For $\beta \gg 1$ we can approximate $\gamma(\beta) \approx 1/W(2\beta)$ and $W(2\beta) \approx \log \beta - \log \log\beta$. Hence, the leading term proportional to $\log^2 \beta$ cancels and we find
\begin{equation}
 \sigma_\mathrm{T}^\text{att.} m_\phi^2 \approx 2 \pi \log \beta (\log \log \beta + 1) \; .
\end{equation}
We emphasize that this asymptotic behavior differs from the one found previously for the classical limit ($\sigma_\mathrm{T}^\text{att.} \propto \log^2 \beta$, cf.\ Appendix~\ref{app:review}) and that in contrast to previous calculations it requires no empirical constant of proportionality. Our new result shows better agreement with the numerically calculated cross section and therefore supersedes the existing expressions.

In order to calculate the viscosity cross section, we follow an analogous calculation to the one above to find
\begin{align}
\sigma_\mathrm{V}^\text{att.} m_\phi^2 \approx \frac{\pi}{2\kappa^2} \ell_{\text{max}}^2 \approx \frac{\pi}{2} \left(1 +\log\beta - \frac{1}{2\,\log\beta}\right)^2\,,
\end{align}
which agrees well with the numerical result. 

\subsubsection{Repulsive potential}

In the repulsive case, for $\beta > 1$ the equation $U_\mathrm{eff}(R) = 1$ implies $R > 1$. For $\ell/\kappa \ll 1$ it is clear that the second term in Eq.~\eqref{eq:ueff} dominates over the first one in the relevant range of $R$ and we find
\begin{equation}
    R_0 \approx W(2\beta) \; .
\end{equation}
For $\ell/\kappa \gg 1$ the first term dominates and we find
\begin{equation}
    R_0 \approx \frac{\ell}{\kappa} \; .
\end{equation}
As before, we use $\ell_\text{max}$ to denote the transition between the two solutions, which occurs when $W(2\beta) \approx R_0$, leading to
\begin{equation}
 \ell_{\text{max}} \approx \kappa W(2\beta) \; .
\end{equation}

For $\ell > \ell_{\text{max}}$ the phase shifts are very small and their contribution to the momentum transfer cross section can be neglected. For $\ell < \ell_{\text{max}}$ we can use a similar calculation as in Appendix~\ref{app:phase_shifts} to show that
\begin{equation}
    \delta'(\ell) \approx \frac{\pi}{2} + \frac{\ell}{\kappa} \int_{R_0}^\infty \frac{\mathrm{d}R}{R^2} \approx \frac{\pi}{2} - \frac{\ell}{\ell_\text{max}} \; .
\end{equation}

Substituting this expression into the formula for the momentum transfer cross section, we find
\begin{equation}
\sigma_\mathrm{T}^\text{rep.} m_\phi^2 \approx \pi \lambda_\mathrm{T} W(2\beta)^2\, ,
\end{equation}
with $\lambda_\mathrm{T} \equiv (1 + \cos 2 + 2 \sin 2)/2 \approx 1.20$. 
For large $\beta$ this expression becomes approximately equal to
\begin{equation}
\sigma_\mathrm{T}^\text{rep.} m_\phi^2 \approx \pi \lambda_\mathrm{T} (\log 2\beta - \log \log 2 \beta)^2 \; ,
\end{equation}
which differs from the result in the literature by the factor $\lambda_T$. We have confirmed that our new result improves the agreement with the numerically calculated cross section and therefore supersedes existing expressions.

For the viscosity cross section we find that for $\ell < \ell_\text{max}$
\begin{align}
    2 \delta'(\ell) \approx \pi - \frac{2\ell}{\ell_{\text{max}}} \; .
\end{align}
Since this expression is close to $\pi$ for small $\ell$, the resulting contribution to the viscosity cross section is suppressed and we need to extend the integration limit beyond $\ell_\text{max}$ to better approximate the cross section. For $\ell > \ell_\text{max}$ we can use a calculation analogous to the case of small $\beta$ to show that
\begin{align}
    2 \delta'(\ell) \approx 2\beta K_1\left(\frac{\ell}{\kappa}\right) \; .
\end{align}
Combining these two expressions yields
\begin{equation}
    \sigma_\mathrm{V}^\text{rep.} \approx \frac{2 \pi}{\kappa^2m_\phi^2}\left[\int_0^{\ell_\text{max}} \ell \sin^2\left(\pi - \frac{2\ell}{\ell_\text{max}}\right) \mathrm{d}\ell  + \int_{\ell_\text{max}}^\infty \ell \sin^2 \left(2\beta K_1\left(\frac{\ell}{\kappa}\right)\right) \mathrm{d}\ell \right] \; .
\end{equation}
The first term can be integrated to yield $\pi \lambda_\mathrm{V} \, W(2\beta)^2$ with $\lambda_\mathrm{V} = (9 - \cos 4 - 4 \sin 4)/16 \approx 0.793$. 
For the second term we note that the argument of the sine is large for $\ell < \kappa \log 2 \beta$ and quickly becomes negligible for larger $\ell$.
Hence, we obtain
\begin{align}
\sigma_\mathrm{V}^\text{rep.} m_\phi^2 & \approx \pi \lambda_\mathrm{V} \, W(2\beta)^2 + \frac{\pi}{2}\left(\log^2 2 \beta - W(2\beta)^2\right) \\
& \approx \pi \log 2\beta \left(\lambda_\mathrm{V} \log 2\beta - (2 \lambda_V - 1) \log \log 2 \beta \right) \; .
\end{align}

To conclude this discussion, we note that for large $\beta$ the scattering of identical particles becomes trivial, as we always find $\sigma_\mathrm{V}^\text{even} \approx \sigma_\mathrm{V}^\text{odd} \approx \sigma_\mathrm{V}$.

\subsection{Intermediate regime}

So far we have focused on the two limiting cases $\beta \ll 1$ and $\beta \gg 1$, where it is possible to derive approximate analytical expressions for the momentum transfer and viscosity cross sections. In the intermediate regime, however, the approximations made above do not apply. Although it is still possible to derive approximate expressions for the phase shifts for $\ell \ll \kappa$ and $\ell \gg \kappa$, the final cross sections turn out to depend sensitively on contributions with $\ell \approx \kappa$.

In particular, for $\beta \gtrsim 0.2$ the numerical results for the momentum transfer cross section differ between the attractive potential and repulsive potential, which cannot be reproduced using the approximations made above for $\beta \ll 1$. Conversely, for $\beta \lesssim 50$ the momentum transfer cross section for an attractive potential exhibits a number of oscillations, which cannot be reproduced in our treatment of the large $\beta$ regime.\footnote{Qualitatively, these oscillations can be understood in terms of the maximal value of $|\delta'(\ell)|$. If
\mbox{$|\delta'(\ell)|_{\text{max}} \approx n \pi$}
for  some integer $n$ the momentum transfer cross section exhibits a dip, whereas for
$|\delta'(\ell)|_{\text{max}} \approx \left(n - \tfrac{1}{2}\right) \pi$ the momentum transfer cross section exhibits a peak. However, we note that these peaks and dips are smeared out when averaging over a range of velocities (see Sec.~\ref{sec:pheno}) and therefore lose relevance in realistic applications.}
Rather than attempting to refine our treatment to reproduce these effects in the intermediate regime, we derive an improved empirical fitting formula that can be used for $0.2 < \beta < 50$.

Specifically, we numerically calculate the momentum transfer cross section for $\beta = 1$ and $\kappa \gg 1$, which is found to be $\sigma_\mathrm{T}^\text{att.} m_\phi^2 / \pi = 2.8$ ($\sigma_\mathrm{T}^\text{rep.} m_\phi^2 / \pi = 1.1$) for the attractive (repulsive) case. We then assume that $\sigma_\mathrm{T} m_\phi^2 / \pi$ can be approximated by $a \log(\beta + b)$ for $1 < \beta < 50$ and determine $a$ and $b$ such that the numerical result at $\beta = 1$ is reproduced and the function connects continuously to the analytical expressions for large $\beta$ at $\beta = 50$. This approach yields
\begin{align}
 \sigma_\mathrm{T}^\text{att.} & = 4.7 \log(\beta + 0.82) \frac{\pi}{m_\phi^2}   \qquad (1 < \beta < 50) \, ,\\
 \sigma_\mathrm{T}^\text{rep.} & = 2.9 \log(\beta + 0.47)\frac{\pi}{m_\phi^2} \qquad (1 < \beta < 50) \; .
\end{align}

For $0.2 < \beta < 1$ the dependence of the momentum transfer cross section on $\kappa$ is non-negligible. To ensure continuity at $\beta = 0.2$, we take the analytical expression obtained for small $\beta$ and multiply by $e^{c(\beta - 0.2)}$, where $c$ is determined such that the momentum transfer cross section is continuous at $\beta = 1$ for $\kappa \gg 1$. This approach yields
\begin{align}
 \sigma_\mathrm{T}^\text{att.} & = \frac{2 \pi \beta^2}{m_\phi^2} \zeta_{1/2}\left(\kappa, \beta\right) e^{0.64(\beta - 0.2)}    \qquad (0.2 < \beta \leq 1) \, ,\\
 \sigma_\mathrm{T}^\text{rep.} & = \frac{2 \pi \beta^2}{m_\phi^2}\zeta_{1/2}\left(\kappa, \beta\right) e^{-0.53 (\beta - 0.2)}    \qquad (0.2 < \beta \leq 1) \; .
\end{align}
Note that for small $\kappa$ this approach leads to a slight discontinuity at $\beta = 1$, which however has no practical consequences.

We follow the exact same approach for the viscosity cross section, with the only difference being that the intermediate regime is shifted to slightly smaller values of $\beta$. We numerically evaluate the cross sections at $\beta = 0.5$, which yields $\sigma_\mathrm{V}^\text{att.} m_\phi^2 / \pi = 1.1$ ($\sigma_\mathrm{V}^\text{rep.} m_\phi^2 / \pi = 0.73$) for the attractive (repulsive) potential, leading to
\begin{align}
 \sigma_\mathrm{V}^\text{att.} & = 2.5 \log(\beta + 1.05) \frac{\pi}{m_\phi^2}   \qquad (0.5 < \beta < 25) \, ,\\
 \sigma_\mathrm{V}^\text{rep.} & = 2.8 \log(\beta + 0.80)\frac{\pi}{m_\phi^2} \qquad (0.5 < \beta < 25) \; ,
\end{align}
and
\begin{align}
 \sigma_\mathrm{V}^\text{att.} & = \frac{4\pi\beta^2}{m_\phi^2} \zeta_n\left(\kappa, 2\beta\right)e^{0.67(\beta - 0.1)}   \qquad (0.1 < \beta \leq 0.5) \, ,\\
 \sigma_\mathrm{V}^\text{rep.} & = \frac{4\pi\beta^2}{m_\phi^2} \zeta_n\left(\kappa, 2\beta\right)e^{-0.37(\beta - 0.1)}   \qquad (0.1 < \beta \leq 0.5) \; ,
\end{align}
where $n = 1$ for distinguishable particles and $n = \tfrac{1}{2}$ ($n = \tfrac{3}{2}$) for identical particles with even (odd) spatial wave function.

\subsection{Summary}
\label{sec:Summary}
In summary, we obtain the following expressions for the momentum transfer cross section $\sigma_\mathrm{T}$ for the scattering of distinguishable particles
 in an attractive or repulsive Yukawa potential:
\begin{align}
\sigma_\mathrm{T}^{\mathrm{att.}} &= \frac{\pi}{m_\phi^2} \times \begin{cases}
2\beta^2 \zeta_{1/2}\left(\kappa, \beta\right) & \beta\leq 0.2 \,, \\
\hspace{7cm}\  & \  \\[-4mm]
2\beta^2 \zeta_{1/2}\left(\kappa, \beta\right) e^{0.64(\beta - 0.2)} & 0.2 < \beta \leq 1\,, \\
\hspace{7cm}\  & \  \\[-4mm]
4.7 \log(\beta + 0.82)  & 1 < \beta < 50\,, \\ 
\hspace{7cm}\  & \  \\[-4mm]
2 \log \beta (\log \log \beta + 1) & \beta \geq 50\;,
\end{cases} 
\label{eq:sigmaT_att_an}
\\
\nonumber\\
\sigma_\mathrm{T}^{\mathrm{rep.}} & = \frac{\pi}{m_\phi^2} \times \begin{cases}
2\beta^2\zeta_{1/2}\left(\kappa, \beta\right) & \beta\leq0.2\,,\\
\hspace{7cm}\  & \ \\[-4mm]
2\beta^2\zeta_{1/2}\left(\kappa, \beta\right) e^{-0.53(\beta - 0.2)} & 0.2 < \beta \leq 1\,, \\
\hspace{7cm}\  & \ \\[-4mm]
2.9 \log(\beta + 0.47) & 1 < \beta < 50\,, \\
\hspace{7cm}\  & \ \\[-4mm]
\lambda_\mathrm{T} (\log 2\beta - \log \log 2 \beta)^2 & \beta \geq 50\;,
\end{cases} 
\label{eq:sigmaT_rep_an}
\end{align}
with
\begin{align}
 \zeta_n(\kappa,\beta) & =\frac{\text{max}(n,\beta\kappa)^2 - n^2}{2 \kappa^2 \beta^2} + \eta\left(\frac{\text{max}(n,\beta\kappa)}{\kappa}\right)\,,\\
 \eta(x) & = x^2 \left[ -K_1\left(x\right)^2 + K_0\left(x\right) K_2\left(x\right)\right] \\
 & \approx - 2 \log \left(\frac{x}{2}\right) - 1 - 2 \gamma_\mathrm{E} + x^2\left[1-\gamma_\mathrm{E}- \log \left(\frac{x}{2}\right)\right] \; .
\end{align}

For the viscosity cross section we obtain
\begin{align}
\sigma_\mathrm{V}^{\mathrm{att.}} & = \frac{\pi}{m_\phi^2}  \times \begin{cases}
4\beta^2 \zeta_n\left(\kappa, 2\beta\right) & \beta\leq0.1\,,\\
\hspace{7cm}\  & \ \\[-4mm]
4\beta^2 \zeta_n\left(\kappa, 2\beta\right) e^{0.67(\beta - 0.1)} & 0.1 < \beta \leq0.5\,,\\
\hspace{7cm}\  & \ \\[-4mm]
2.5 \log(\beta + 1.05) & 0.5 < \beta < 25\,, \\
\hspace{7cm}\  & \ \\[-4mm]
\frac{1}{2} \left(1 + \log{\beta} - \frac{1}{2\log\beta}\right)^2 & \beta\geq25 \;,
\end{cases} 
\label{eq:sigmaV_att_an}
\\
\nonumber\\
\sigma_\mathrm{V}^{\mathrm{rep.}} & = \frac{\pi}{m_\phi^2} \times \begin{cases}
4\beta^2 \zeta_n\left(\kappa, 2\beta\right) & \beta\leq0.2\,,\\
\hspace{7cm}\  & \ \\[-4mm]
4\beta^2 \zeta_n\left(\kappa, 2\beta\right) e^{-0.37(\beta - 0.1)} & 0.1 < \beta \leq0.5\,,\\
\hspace{7cm}\  & \ \\[-4mm]
2.8 \log(\beta + 0.80) & 0.5 < \beta < 25\,, \\
\hspace{7cm}\  & \ \\[-4mm]
\log 2\beta \left(\lambda_\mathrm{V} \log 2\beta - (2 \lambda_V - 1) \log \log 2 \beta \right) & \beta \geq 25\;,
\end{cases} 
\label{eq:sigmaV_rep_an}
\end{align}
with
\begin{align}
 \lambda_\mathrm{T} & = (1 + \cos 2 + 2 \sin 2)/2 \, , \\
 \lambda_\mathrm{V} & = (9 - \cos 4 - 4 \sin 4)/16 \; ,
\end{align}
where $n = 1$ for distinguishable particles and $n = \tfrac{1}{2}$ ($n = \tfrac{3}{2}$) for identical particles with even (odd) spatial wave function.

\section{Results}
\label{sec:results}
\subsection{Comparison with numerical results}

\begin{figure}
    \centering
    \includegraphics[width=\textwidth]{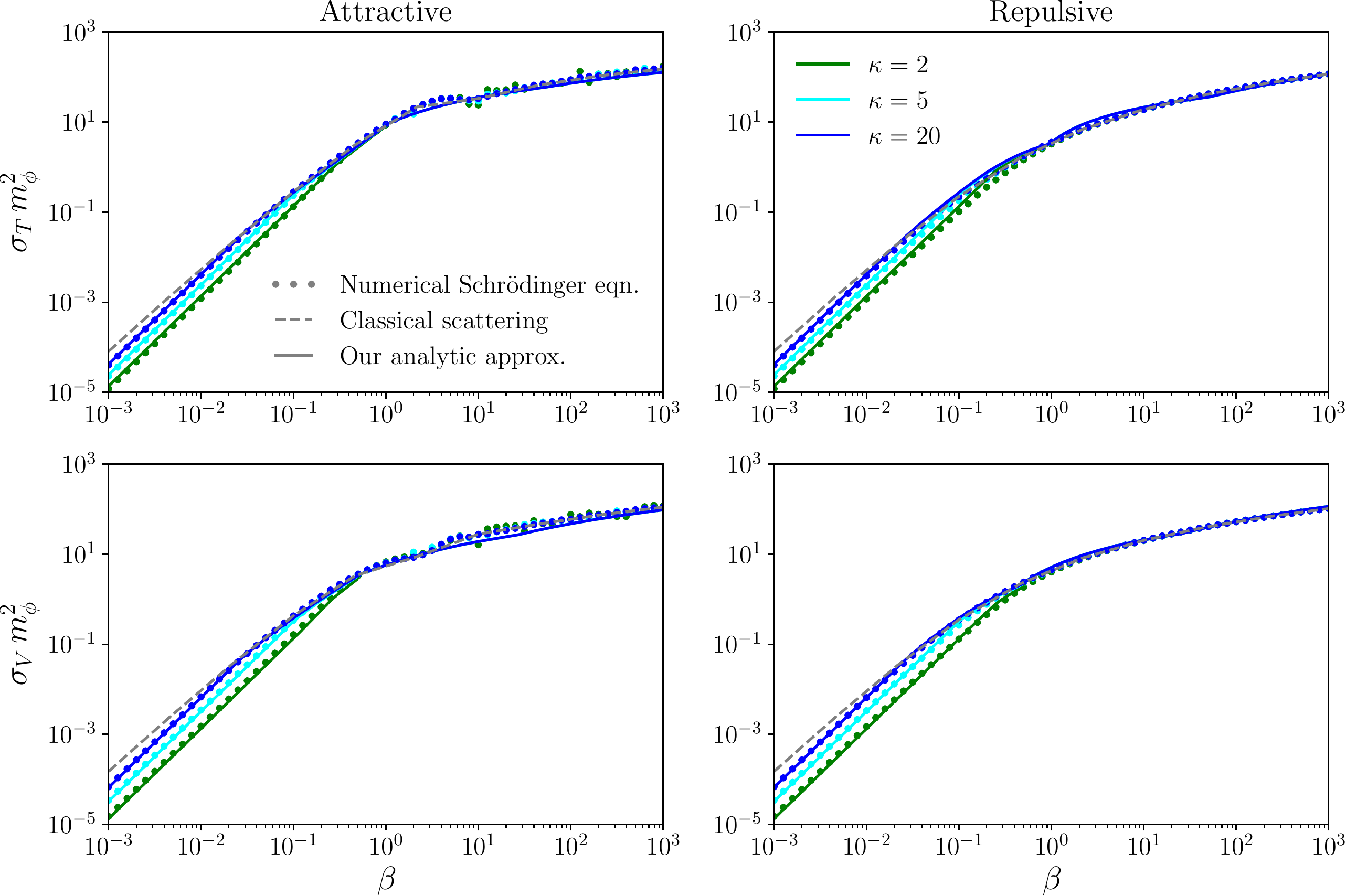}
    \caption{Dimensionless cross sections $\sigmaT m_\phi^2$ and $\sigmaV m_\phi^2$ as functions of $\beta$, for attractive and repulsive Yukawa potentials and for various values of $\kappa$. 
    Solid lines show our analytic results, derived in Sec.~\ref{sec:analytic}, which are in good agreement with exact numerical results from solving the Schr\"{o}dinger equation (dots). 
    Dashed lines correspond to purely classical scattering.}
    \label{fig:beta}
\end{figure}

We compare our analytic formulas for the semi-classical regime, given in Eqs.~(\ref{eq:sigmaT_att_an}--\ref{eq:sigmaV_rep_an}), to exact results obtained by numerically solving the Schr\"{o}dinger equation.
In Fig.~\ref{fig:beta}, the solid lines show our analytic approximations for the dimensionless cross sections $\sigmaT m_\phi^2$ (top) and $\sigmaV m_\phi^2$ (bottom) as functions of $\beta$ for an attractive (left) and repulsive (right) potential, each for different values of $\kappa$.
These results are in better agreement with the exact numerical results (dots) compared to the classical scattering cross section (dashed line) that is strictly valid in the limit $\kappa \to \infty$.
Outside this limit, however, the cross section is clearly dependent on $\kappa$ in the Coulomb regime ($\beta \lesssim 1$), but independent of $\kappa$ in the strongly-coupled regime ($\beta \gtrsim 1$). 

\begin{figure}
    \centering
    \includegraphics[width=\textwidth]{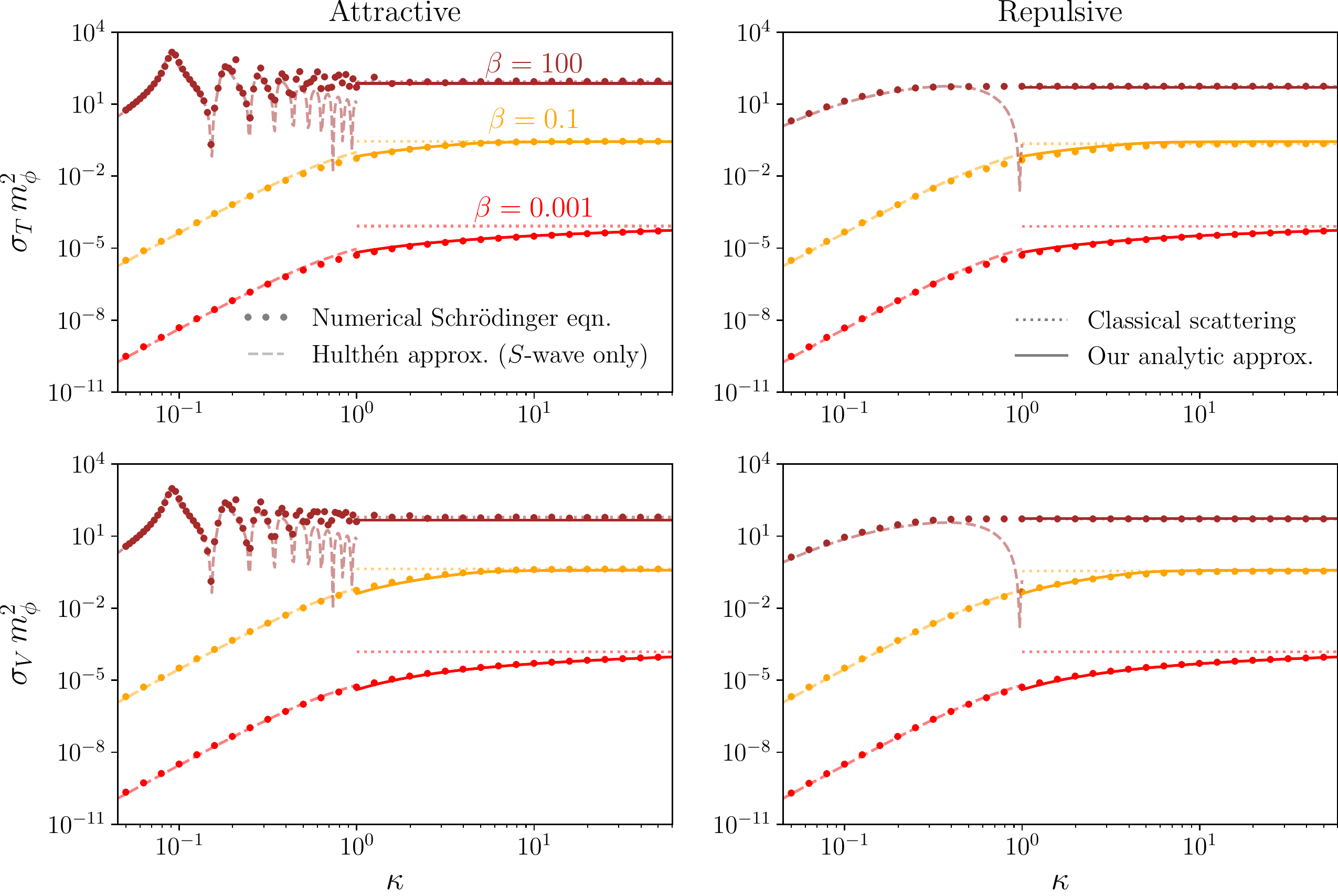}
    \caption{Dimensionless cross sections $\sigmaT m_\phi^2$ and $\sigmaV m_\phi^2$ as functions of $\kappa$, for attractive and repulsive Yukawa potentials and for various values of $\beta$. 
    Solid lines show our analytic results for the semi-classical regime $\kappa > 1$, which are in good agreement with exact numerical results from solving the Schr\"{o}dinger equation (dots). 
    Dotted lines correspond to purely classical scattering.
    Dashed lines represent analytic approximations for scattering in the quantum regime $\kappa  < 1$ using the Hulth\'{e}n potential.
    }
    \label{fig:kappa}
\end{figure}

Fig.~\ref{fig:kappa} further illustrates the $\kappa$ dependence of our results, as well as the complementarity with other treatments for Yukawa scattering.
As above, we show the attractive (left) and repulsive (right) dimensionless cross sections $\sigmaT m_\phi^2$ (top) and $\sigmaV m_\phi^2$ (bottom), now as functions of $\kappa$ and each for several values of $\beta$.
Our exact numerical results from solving the Schr\"{o}dinger equation (dots) span both the semi-classical ($\kappa > 1$) and quantum ($\kappa < 1$) regimes.
For the former, our analytic formulas (solid lines) approximately capture the true $\kappa$ dependence of these cross sections for small values of $\beta$ down to $\kappa \sim 1$, which is neglected for the classical scattering formula (dotted lines).
On the other hand, the quantum regime is typically dominated by $S$-wave scattering which can be approximated using the Hulth\'{e}n potential (dashed lines), including the effect of resonances and anti-resonances~\cite{Tulin:2013teo}.\footnote{In this approximation, one replaces the Yukawa potential by the Hulth\'{e}n potential $U(r) = \pm \frac{\alpha_\chi \delta e^{-\delta r}}{1 - e^{-\delta r}  }$, which shares a similar Coulomb-like behavior $\sim 1/r$ at small $r$ and is screened as $e^{-\delta r}$ at large $r$, with screening parameter $\delta \sim \mathcal{O}(m_\phi)$~\cite{Cassel:2009wt}.
The $S$-wave Schr\"{o}dinger equation can be solved exactly in terms of hypergeometric functions, which yields the phase shift 
\begin{equation}
\delta_0 = \arg\left( \frac{ i \Gamma(\lambda_+ + \lambda_- -2)}{\Gamma(\lambda_+) \Gamma(\lambda_-)} \right) \, ,
\qquad
\lambda_\pm = 1 + \frac{i \kappa}{n} \pm \frac{\kappa}{n} \times \left\{ \begin{array}{cc} i \sqrt{ 2\beta n + 1} & {\rm repulsive} \\ \sqrt{2 \beta n - 1} & {\rm attractive} \end{array} \right. \, ,
\nonumber\end{equation}
where $\Gamma$ is the usual gamma function and $n = \delta/m_\phi \approx 1.6$ is a numerical constant~\cite{Tulin:2013teo}.\vspace{2cm}}
Across the two regimes, our new semi-classical formulas combined with the Hulth\'{e}n potential approximation provide nearly complete analytic coverage of the entire parameter space for Yukawa scattering, with the exception of the region $\kappa \sim 1$ where both approximations break down. In practice, a good approximation for $\kappa \approx 1$ can be obtained by simply interpolating between the two different regimes.

\begin{figure}
    \centering
    \includegraphics[width=\textwidth]{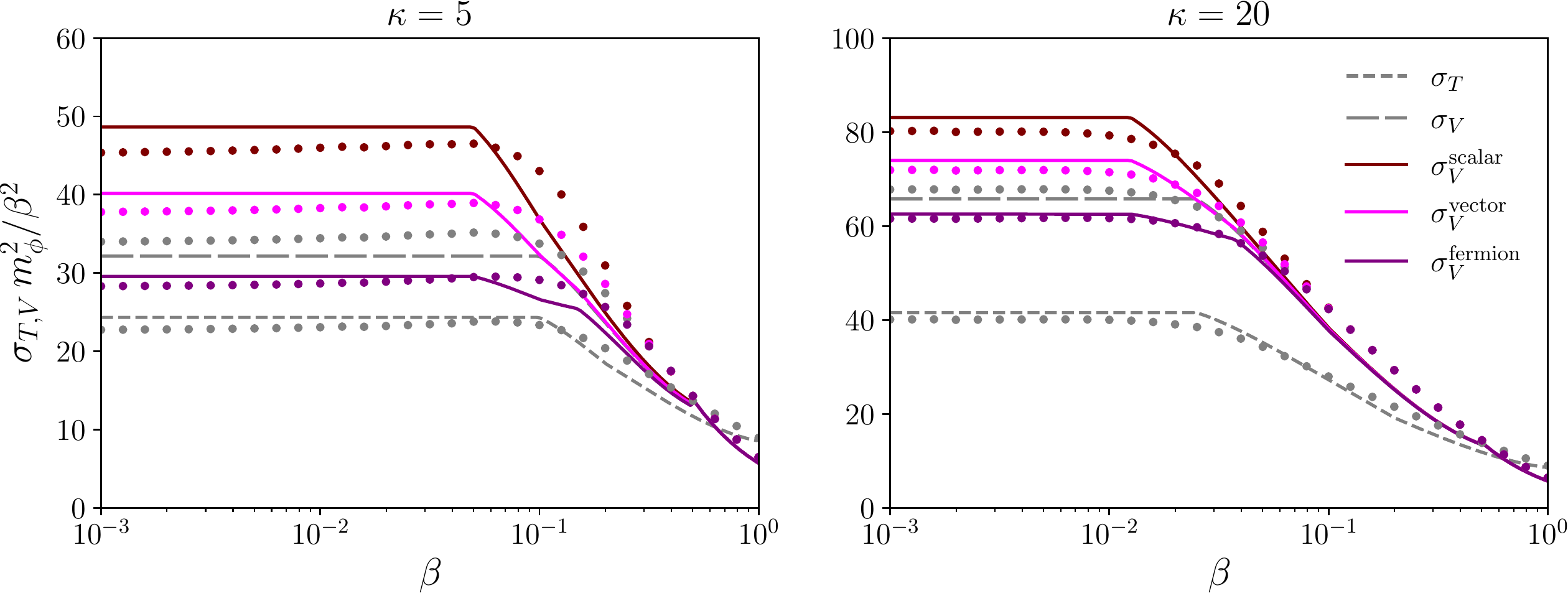}
    \caption{Cross sections for distinguishable vs. identical particles for an attractive Yukawa potential, with $\kappa=5$ (left) and $\kappa=20$ (right).
    Distinguishable cross sections $\sigmaT$ (short-dashed) and $\sigmaV$ (long-dashed) are indicated in gray, while colored curves (solid) correspond to identical scattering, depending on the DM particle spin $s=0,\tfrac{1}{2},1$.
    Continuous lines are our analytic results and dots are numerical results from solving the Schr\"{o}dinger equation.}
    \label{fig:ident}
\end{figure}

Finally, we consider in Fig.~\ref{fig:ident} the scattering of identical particles. To highlight the differences between the various cases, we focus on $\beta < 1$ and plot the cross sections multiplied by $1/\beta^2$, which approaches a constant for $\beta \to 0$. The largest differences are found for small $\kappa$ (left panel), where only a small number of phase shifts contribute to the cross section. 
Since $\sigma_\mathrm{V}^\text{even} > \sigma_\mathrm{V}^\text{odd}$ the cross section is largest for $\sigma_\mathrm{V}^\text{scalar}$ (where only the even part contributes) and smallest for $\sigma_\mathrm{V}^\text{fermion}$ (where the odd part dominates). 
For $\kappa = 5$, using the momentum transfer cross section instead of the appropriate viscosity cross section for identical particles can lead to a difference as large as a factor of 2.

\subsection{Velocity averaging}
\label{sec:averaging}

Let us now apply our results above to the description of DM self-scattering in astrophysical systems. For this purpose we need to account for the distribution of DM velocities in such systems. Here we assume that the DM velocity distribution is approximately isothermal, i.e.\ given by a Maxwell-Boltzmann (MB) distribution with one-dimensional velocity dispersion $v_0$ independent of radius. The relative velocity $v_\text{rel}$ between two DM particles then follows a MB distribution with velocity dispersion $\sqrt{2} v_0$, such that the typical relative velocity is
\begin{equation}
    \langle v_\text{rel} \rangle = \frac{4}{\sqrt{\pi}} v_0 \; ,
\end{equation}
where the brackets denote the expectation value with respect to the MB distribution.

The expected rate $\Gamma$ of DM self-scattering is then proportional to $\langle \sigma v_\text{rel} \rangle$. Here we are interested not in the rate of scattering, but in the rate at which DM self-interactions transfer momentum or energy. In the literature, this is commonly done by simply replacing $\sigma$ by either $\sigma_\mathrm{T}$ or $\sigma_\mathrm{V}$ and calculating $\langle \sigma_\mathrm{T,V} v_\text{rel} \rangle$. However, as shown in Appendix~\ref{app:averaging}, the momentum and energy transfer rates are in fact proportional to $\langle \sigma_\mathrm{T} v_\text{rel}^2 \rangle$ and $\langle \sigma_\mathrm{V} v_\text{rel}^3 \rangle$, respectively. 
By defining
\begin{align}
\overline{\sigma_\mathrm{T}} = \frac{\langle \sigma_\mathrm{T} v_\text{rel}^2\rangle}{16 \sqrt{2} v_0^2 / \pi} \, , \qquad 
\overline{\sigma_\mathrm{V}} = \frac{\langle \sigma_\mathrm{V} v_\text{rel}^3 \rangle}{24 / \sqrt{\pi} v_0^3} \; ,
\end{align}
we can write these rates as
\begin{align}
\Gamma_p \equiv \frac{\langle \dot{p} \rangle}{\langle p \rangle} = \frac{\rho}{m_\chi} \, \overline{\sigma_\mathrm{T}} \, \langle v_\text{rel} \rangle \, , \qquad
\Gamma_E \equiv \frac{\langle \dot{E_\perp} \rangle}{\langle E \rangle} = \frac{\rho}{m_\chi} \, \overline{\sigma_\mathrm{V}} \, \langle v_\text{rel} \rangle \; .
\end{align}
These rates govern the evolution of DM halos under the influence of DM self-interactions. For example, we expect core formation to occur whenever $\Gamma_E \, t_0 > 1$, where $t_0$ is the age of the system~\cite{Kaplinghat:2015aga,Sokolenko:2018noz,Robertson:2020pxj}.

The rate of momentum and energy transfer place more weight on collisions with larger velocity than the scattering rate, while the effect of collisions with smaller velocity is suppressed. This implies in particular that a finite result is obtained for a cross section that scales proportional to $v_\text{rel}^{-4}$, which is not the case for the naive average $\langle \sigma_\mathrm{T,V} v_\text{rel} \rangle$. For a velocity-independent cross section, on the other hand, one finds $\overline{\sigma_\mathrm{T}} \approx 0.83 \sigma_\mathrm{T}$ and $\overline{\sigma_\mathrm{V}} \approx 1.34 \sigma_\mathrm{V}$.

\begin{figure}
    \centering
    \includegraphics[width=\textwidth]{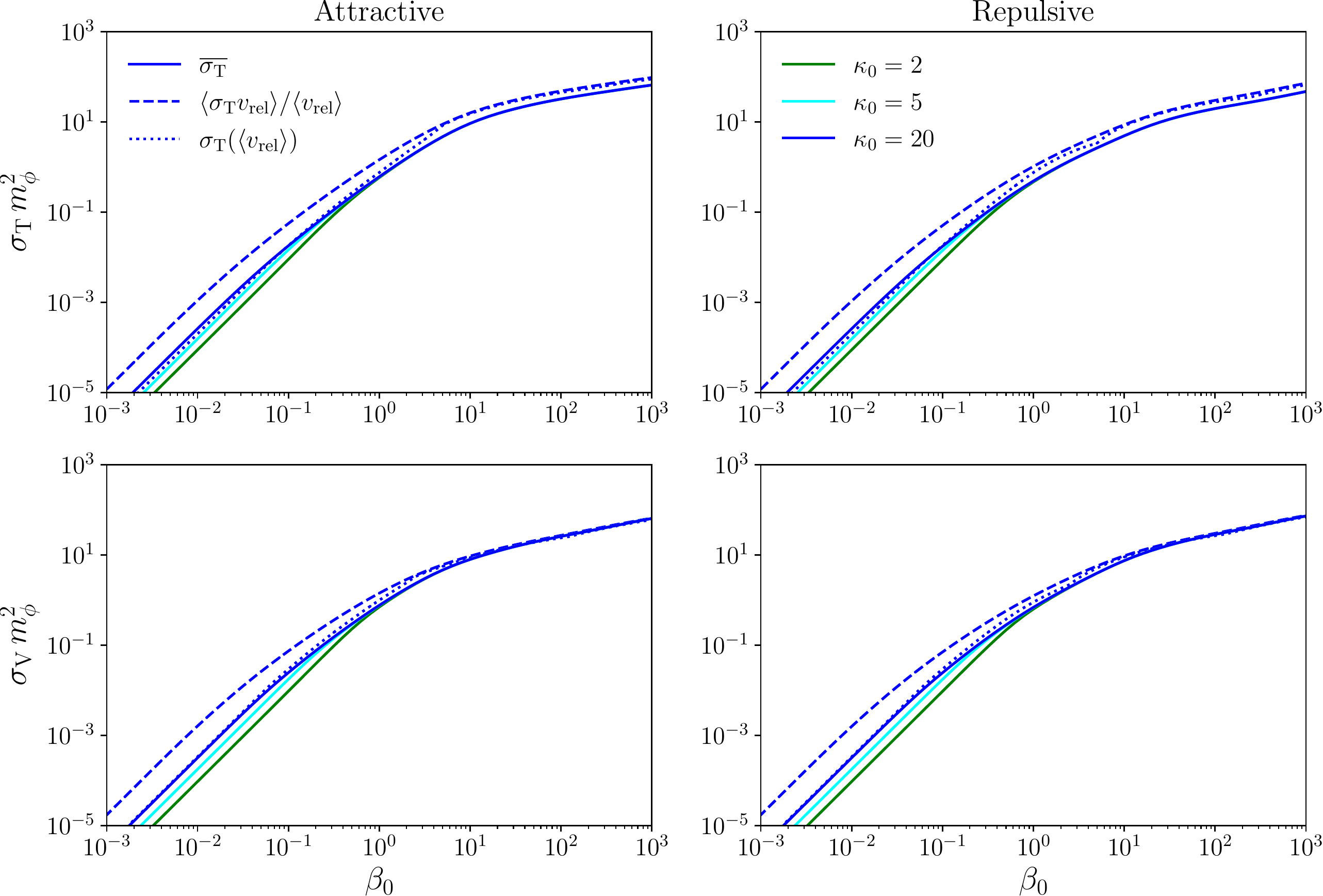}
    \caption{Averaged momentum transfer cross section (top) and viscosity cross section (bottom) for the attractive (left) and repulsive (right) Yukawa potential following different averaging prescriptions. Solid lines correspond to the averaging proposed in this work, while dashed and dotted lines indicate alternative approaches as explained in the text. The different colors illustrate the dependence on $\kappa_0$.}
    \label{fig:sigma_averaging}
\end{figure}

As shown in Appendix~\ref{app:averaging}, in the semi-classical regime $m_\phi^2 \overline{\sigma_\mathrm{T}}$ and $m_\phi^2 \overline{\sigma_\mathrm{V}}$ are functions of only $\beta_0$ and $\kappa_0$, which are defined as the values of $\beta$ and $\kappa$ obtained for $v = v_0$. We show these functions in Fig.~\ref{fig:sigma_averaging} and compare to existing approaches where $\overline{\sigma_\mathrm{T}}$ is obtained either by calculating $\langle \sigma_\mathrm{T} v_\text{rel} \rangle/\langle v_\text{rel} \rangle$ or simply by evaluating $\sigma_\mathrm{T}$ at $v = \langle v_\text{rel} \rangle$. We point out that the wiggles for intermediate $\beta$ disappear when the averaging is performed and that our analytical expressions give very similar results to the full numerical expressions. 

\subsection{Phenomenological implications}
\label{sec:pheno}

In practice we are interested in astrophysical systems ranging from dwarf galaxies with $v_0 \sim 20\,\mathrm{km\,s^{-1}}$ to galaxy clusters with $v_0 \sim 1000\,\mathrm{km\,s^{-1}}$. For all of these systems to fall into the semi-classical regime, we require $\kappa_\text{min} \equiv \kappa(v_\text{min}) > 1$, which implies $m_\chi v_\text{min} > 2 m_\phi$ and therefore
\begin{equation}
\frac{m_\phi}{m_\chi} \lesssim 10^{-4} \; .  \label{eq:ineq}
\end{equation}
In other words, the semi-classical regime requires a large hierarchy between the mediator mass and the DM mass.

As we have seen above, for $\beta < 1$ the momentum transfer and viscosity cross sections exhibit a strong velocity dependence: $\sigma_\mathrm{T,V} \propto \beta^2 \propto v^{-4}$. Given typical upper bounds at dwarf galaxy scales of $\sigma_\mathrm{T} / m_\chi \lesssim 50\,\mathrm{cm^2 \, g^{-1}}$ we then predict unobservably small effects from DM self-interactions in larger astrophysical systems like groups or clusters. The much more interesting case is therefore that $\beta > 1$ in the astrophysically interesting range, such that the velocity dependence is much weaker and $\sigma_\mathrm{T,V} / m_\chi$ only changes by one or two orders of magnitude between dwarf and cluster scales. 

To get a rough sense of scale we note that for $\beta \gg 1$ we can approximately write
\begin{equation}
m_\phi^2 \, \sigma_\mathrm{T} \approx c\, , \label{eq:simple_approx}
\end{equation}
where $c$ is a number of the order of $100$ that depends only logarithmically on $\beta$. It then follows that in order to obtain a momentum transfer cross section in the observationally interesting range ($\sigma_\mathrm{T} / m_\chi \sim 1 \, \mathrm{cm^2 \, g^{-1}}$) one must require $m_\chi m_\phi^2 \sim 0.02 \, \mathrm{GeV^3}$. Substituting this expression into Eq.~(\ref{eq:ineq}), we obtain $m_\phi \lesssim 10\,\mathrm{MeV}$. We conclude that rather light mediators are necessary in order to obtain phenomenologically interesting cross sections in the semi-classical regime.\footnote{Note that such light mediators cannot be in thermal equilibrium with the Standard Model, because they would spoil the agreement between the observed element abundances and predictions from primordial nucleosynthesis. These constraints are evaded if the dark sector is (somewhat) colder than the Standard Model. In such a setup, the DM abundance can be set, for example, via the freeze-in mechanism~\cite{Bernal:2015ova}.}

\begin{table}[t]
    \caption{Velocity averaged self-interaction cross sections for $m_\chi = 190\,\mathrm{GeV}$, $m_\phi = 3 \, \mathrm{MeV}$, $\alpha_\chi = 0.5$ at different astrophysical scales.\label{tab:CS}}
    \centering
    \begin{tabular}{>{}p{0.15\textwidth}>{\centering\arraybackslash}p{0.1\textwidth}>{\centering\arraybackslash}p{0.1\textwidth}>{\centering\arraybackslash}p{0.1\textwidth}>{\centering\arraybackslash}p{0.1\textwidth}>{\centering\arraybackslash}p{0.1\textwidth}>{\centering\arraybackslash}p{0.1\textwidth}>{\centering\arraybackslash}p{0.1\textwidth}}
    \hline
    \hline
System & $\langle v \rangle$ & $\beta_0$ & $\kappa_0$ & $\overline{\sigma_\mathrm{T}^\text{att.}}/m_\chi$ & $\overline{\sigma_\mathrm{T}^\text{rep.}}/m_\chi$ & $\overline{\sigma_\mathrm{V}^\text{att.}}/m_\chi$ & $\overline{\sigma_\mathrm{V}^\text{rep.}}/m_\chi$ \\
 & & & & $[\mathrm{cm^2 \, g^{-1}}]$ & $[\mathrm{cm^2 \, g^{-1}}]$ & $[\mathrm{cm^2 \, g^{-1}}]$ & $[\mathrm{cm^2 \, g^{-1}}]$ \\
     \hline
Dwarf galaxy & 50 & 2890 & 2.34 & 10.9 & 9.0 & 11.7 & 13.6\\
Galaxy & 250 & 116 & 11.7 & 4.3 & 2.6 & 3.5 & 3.8\\
Galaxy group & 1150 & 5.46 & 53.8 & 0.66 & 0.36 & 0.64 & 0.54\\
Galaxy cluster & 1900 & 2.00 & 88.9 & 0.20 & 0.14 & 0.23 & 0.19 \\     \hline
     \hline
    \end{tabular}
\end{table}

\begin{figure}
    \centering
    \includegraphics[width=0.7\textwidth]{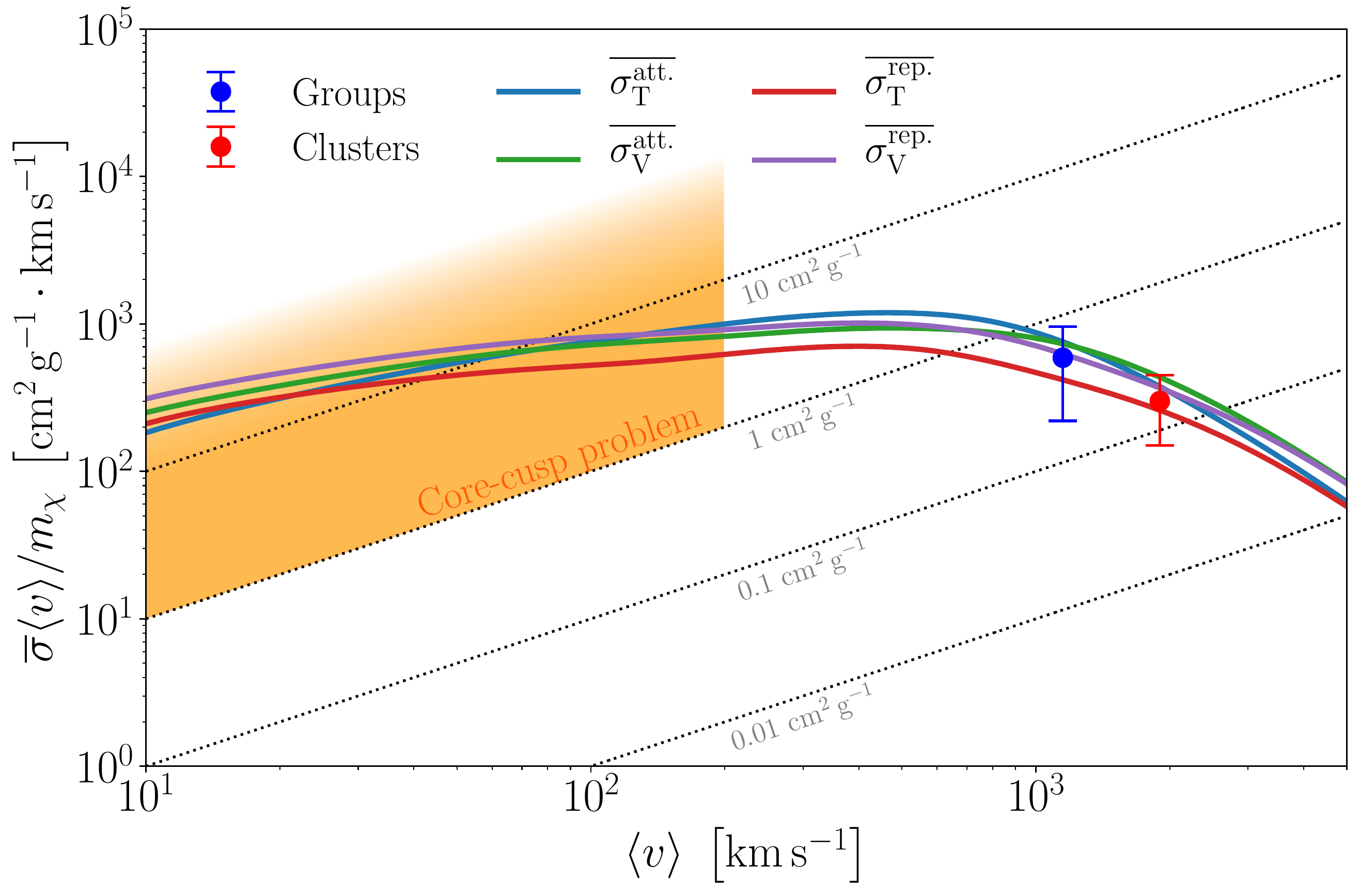}
    \caption{Averaged momentum transfer cross section (blue) and viscosity cross section (green) for $m_\chi = 190\,\mathrm{GeV}$, $m_\phi = 3 \, \mathrm{MeV}$, $\alpha_\chi = 0.5$. For comparison we show $\langle\sigma v\rangle/m_\chi$ inferred from the observation of core sizes from Ref.~\cite{Sagunski:2020spe} and the shaded region indicates the range of cross sections relevant for the core-cusp problem.}
    \label{fig:CS}
\end{figure}

As an example, we provide in Tab.~\ref{tab:CS} and Fig.~\ref{fig:CS} the cross sections $\sigma_\mathrm{T}$ and $\sigma_\mathrm{V}$ as functions of $\langle v_\text{rel} \rangle$ for the benchmark point $m_\chi = 190\,\mathrm{GeV}$, $m_\phi = 3 \, \mathrm{MeV}$, $\alpha_\chi = 0.5$. This benchmark point has been chosen in such a way that the self-interaction cross sections at the scales of groups and clusters agree with what has recently been inferred from the observation of core sizes~\cite{Sagunski:2020spe}.

\begin{figure}
    \centering
    \includegraphics[width=0.49\textwidth,clip,trim=5 0 30 0]{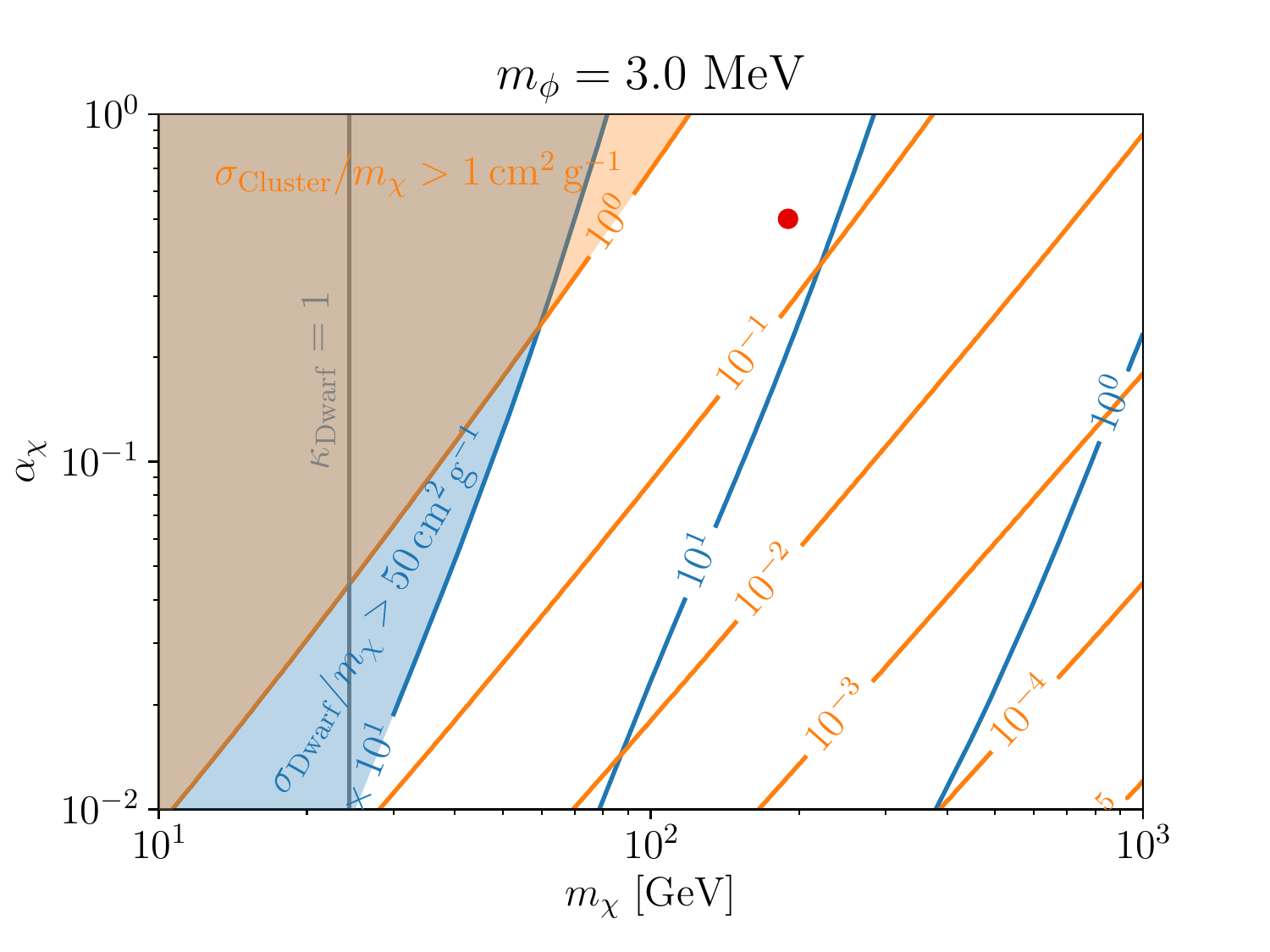}\hfill
    \includegraphics[width=0.49\textwidth,clip,trim=5 0 30 0]{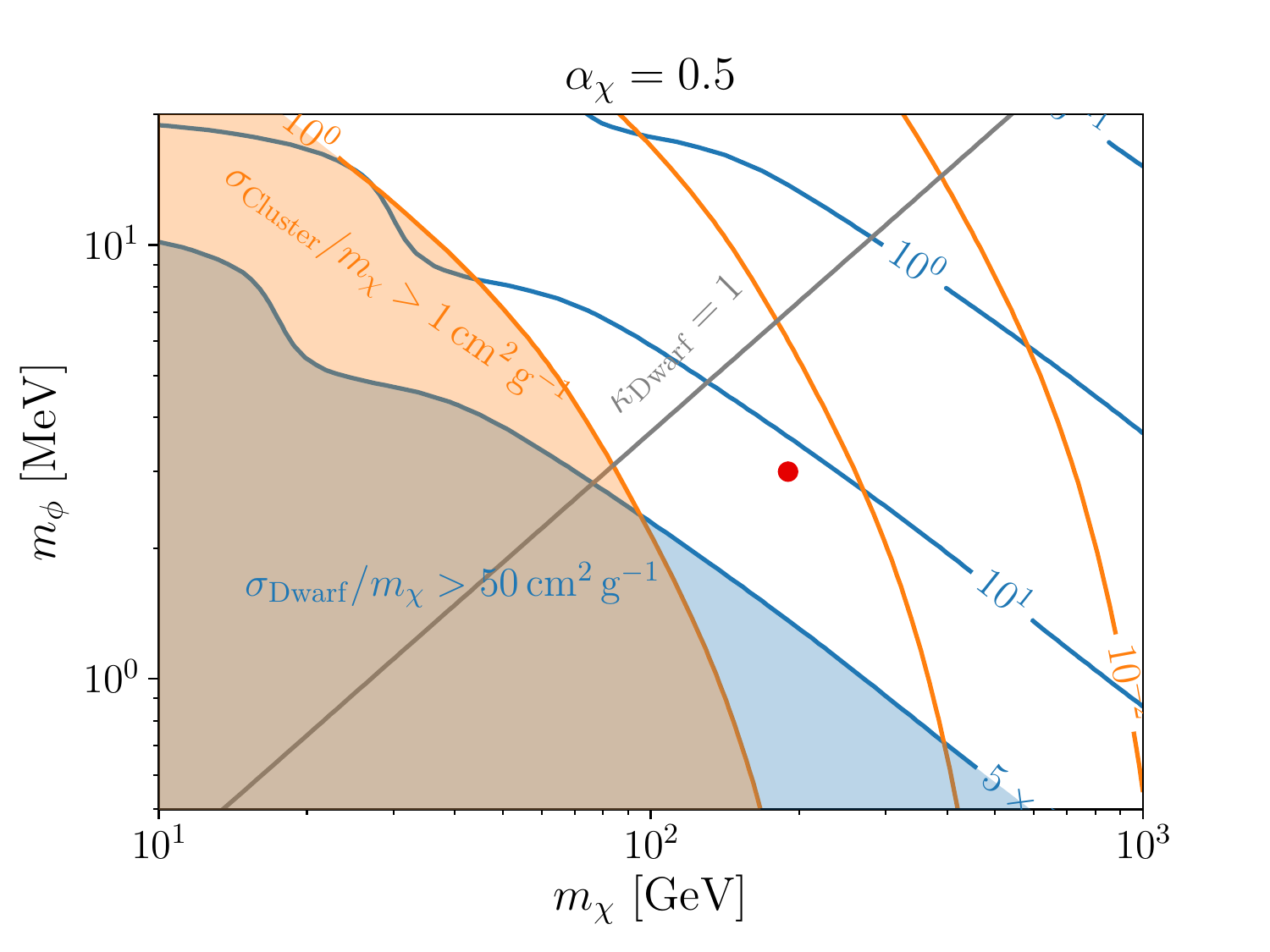}
    \caption{Self-interaction cross section at cluster scales (orange contours) and at dwarf scales (blue contours) as a function of the DM mass $m_\chi$ and the coupling $\alpha_\chi$ (left) and as a function of the DM mass and the mediator mass $m_\phi$ (right). Shaded regions represent approximate bounds from astrophysical observations. The gray line indicates the transition between the quantum regime (left) and the semi-classical regime (right) for dwarf galaxies. At cluster scales self-interactions are always in the semi-classical regime. For concreteness, we consider the viscosity cross section for a repulsive potential, but very similar plots are obtained in the other cases. The red dot indicates the benchmark point considered in Fig.~\ref{fig:CS} and Tab.~\ref{tab:CS}.}
    \label{fig:sigma_pheno}
\end{figure}

To more comprehensively explore the allowed parameter space, we show in Fig.~\ref{fig:sigma_pheno} how the predicted self-interaction cross sections at dwarf and cluster scales depend on the underlying parameters. For concreteness, we consider the viscosity cross section for a repulsive potential, but very similar plots are obtained in the other cases. In the left panel, we fix $m_\phi = 3\,\mathrm{MeV}$ and vary $\alpha_\chi$ and $m_\chi$. We find that on dwarf scales (where $\beta_0 \gg 1$) $\sigma_\mathrm{V}$ depends only very weakly on $\alpha_\chi$ and $m_\chi$, so that the dominant dependence stems from the overall factor $m_\chi^{-1}$. On cluster scales (where $\beta_0 \approx 1$) the dependence of $\sigma_\mathrm{V}$ on the model parameters is more pronounced, such that it is possible to vary the constraints on the two different scales independently. The grey vertical line corresponds to $\kappa_0 = 1$ for $v_0 = 50\,\mathrm{km\,s^{-1}}$ and indicates the transition between the semi-classical regime (to the right) and the quantum regime (to the left). In the quantum regime, we calculate the self-interaction cross section using the Hulth{\'e}n approximation, interpolating linearly between $\kappa = 0.4$ and $\kappa = 1$ to ensure a smooth transition.

In the right panel of Fig.~\ref{fig:sigma_pheno} we fix $\alpha_\chi = 0.5$ and vary $m_\phi$. As expected from Eq.~(\ref{eq:simple_approx}), we find that the cross sections depend strongly on $m_\phi$ and it is not possible to obtain large self-interaction cross sections for $m_\phi > 10\,\mathrm{MeV}$ within the semi-classical regime (to the bottom right of the diagonal grey line). Nevertheless, for smaller mediator masses it is possible to obtain phenomenologically interesting cross sections on both dwarf and cluster scales. The benchmark point considered as an example in Fig.~\ref{fig:CS} and Tab.~\ref{tab:CS} is indicated by the red dot in both panels.

\section{Conclusions \label{sec:concl}} 
Self-interacting DM has long been advocated to alleviate tensions between astrophysical observations and $N$-body simulations of collisionless cold DM. 
For many models, self-interactions are represented by a Yukawa potential and fall within the semi-classical regime, where the de Broglie wavelengths of DM particles are small compared to the characteristic length scales involved in the interaction. 
However, exploring the parameter space of this regime is numerically intensive using the standard partial wave calculation, which requires summing over a large number of $\ell$ modes.
Many previous studies have relied on analytic formulas for the momentum transfer cross section derived for classical scattering of distinguishable particles.
However, this description necessarily breaks down close to the boundary of the semi-classical regime and also for the scattering of identical particles.   

In this work, we provided a fully quantum mechanical description of DM self-interactions in the semi-classical approximation, deriving analytic expressions for the associated cross sections. 
We demonstrated that the momentum transfer cross section $\sigma_\mathrm{T}$ receives relevant quantum corrections for weak potentials, and derived improved asymptotic expressions for strong potentials. 
We also derived, for the first time, analytic expressions for the viscosity cross section, $\sigma_\mathrm{V}$, which captures the rate of energy transfer and turns out to be a more useful quantity than $\sigma_\mathrm{T}$ in the case of identical particles. 
We established the different formulas for $\sigma_\mathrm{V}$ that arise when considering scattering of identical DM particles with spin $s\le 1$. 
These formulas are summarized in Sec.~\ref{sec:Summary}.

To validate our results, we compared them with the cross section obtained by numerically solving the Schr\"{o}dinger equation and found excellent agreement (see Figs.~\ref{fig:beta} and \ref{fig:kappa}). We pointed out that with the addition of our results, almost the entire parameter space of self-interactions in a Yukawa potential can now be described analytically. 
We have implemented our results in the new code CLASSICS, available at \url{https://github.com/kahlhoefer/CLASSICS}.

As an application of our results, we considered the case of DM self-scattering in astrophysical systems. We argued that the velocity-averaged cross section, $\langle \sigma_\mathrm{T,V} v \rangle$ currently used to derive bounds on $\sigma_\mathrm{T,V}$ is imprecise. We showed that the rate of momentum (energy) transfer is in fact proportional to $\langle\sigma_\mathrm{T} v^2 \rangle$ ($\langle\sigma_\mathrm{V} v^3 \rangle$), which can lead to sizeable differences for velocity-dependent scattering. 

Using this updated treatment for velocity averaging and our analytic results for the cross sections, we demonstrated that for sufficiently small mediator masses phenomenologically interesting cross sections at dwarf and cluster scales can be accommodated within the semi-classical regime. It will be exciting to implement these cross sections into numerical simulations in order to study their observational effects in more detail.

An interesting direction for future research is to apply our methodology to models that generate interactions different from the Yukawa potential considered here. Our results suggest that even in situations where it is impossible to calculate the phase shifts analytically, solving the semi-classical integrals numerically may yield an accurate approximation to the full solution of the Schr\"{o}dinger equation. This approach opens up a new avenue for classifying models of self-interacting DM, which is complementary to the recently proposed effective-range formalism applicable to $S$-wave scattering~\cite{Chu:2019awd}.

\section*{Acknowledgements}
We thank Camilo Garcia-Cely, Sophia Gad-Nasr, Andrew Robertson and Kai Schmidt-Hoberg for discussions. This work has been funded by the DFG Emmy Noether Grant No.\ KA 4662/1-1, the DFG grant CRC-TR 211 ``Strong-interaction matter under extreme conditions'' and the Natural Sciences and Engineering Research Council of Canada.

\appendix

\section{Review of existing results for the classical limit}
\label{app:review}

So far, analytic formulas exist only for the momentum transfer cross section $\sigma_\mathrm{T}$ and in the limit $\kappa \to \infty$. In the weak coupling regime $(\beta \ll 1)$, the theory of Coulomb scattering becomes applicable, and $\sigma_\mathrm{T}$ is given by the Coulomb scattering cross section~\cite{FORTOV20051}
\be
\sigma_\mathrm{T}\simeq\frac{2\pi}{m_\phi^2}\beta^2\log(1+\beta^{-2})
\ee
for both attractive and repulsive potentials. In the strong coupling regime $(\beta \gg 1)$, the scattering resembles that of a hard sphere potential and the momentum transfer cross section is a geometric cross section of the form $\sigma_\mathrm{T}\sim \pi R^2$. For an attractive  potential it is approximately given by~\cite{PhysRevLett.90.225002,FORTOV20051}
\be
\sigma_\mathrm{T}\simeq \frac{0.81\pi}{m_\phi^2}(1+\log\beta-(2\log\beta)^{-1})^2\; ,
\ee
whereas for a repulsive potential one finds~\cite{PhysRevE.70.056405}
\be
\sigma_\mathrm{T}\simeq \frac{\pi}{m_\phi^2}(\log2\beta-\log\log2\beta)^2\,.
\ee
In the intermediate regime, only empirical fit formulas for $\sigma_\mathrm{T}$ exist. Following Ref.~\cite{Cyr-Racine:2015ihg}, we can parametrize $\sigma_\mathrm{T}$ for an attractive Yukawa potential as
\begin{equation}
\sigma_\mathrm{T}=\begin{cases}
\frac{2\pi}{m_\phi^2}\beta^2\log(1+\beta^{-2}) & \beta\lesssim10^{-2}\,,\\
\frac{7\pi}{m_\phi^2}\frac{\beta^{1.8}+280(\beta/10)^{10.3}}{1+1.4\beta+0.006\beta^4+160(\beta/10)^{10}} & 10^{-2}\lesssim\beta\lesssim10^2\,,\\
\frac{0.81\pi}{m_\phi^2}(1+\log\beta-(2\log\beta)^{-1})^2 & \beta\gtrsim10^2\;.
\end{cases}
\label{eq:sigmaT_attr}
\end{equation}
For a repulsive Yukawa potential, the corresponding fitting formula is~\cite{Cyr-Racine:2015ihg}
\begin{equation}
\sigma_\mathrm{T}=\begin{cases}
\frac{2\pi}{m_\phi^2}\beta^2\log(1+\beta^{-2}) & \beta\lesssim10^{-2}\,,\\
\frac{8\pi}{m_\phi^2}\frac{\beta^{1.8}}{1+5\beta^{0.9}+0.85\beta^{1.6}} & 10^{-2}\lesssim\beta\lesssim10^4\,,\\
\frac{\pi}{m_\phi^2}(\log2\beta-\log\log2\beta)^2 & \beta\gtrsim10^4\;.
\end{cases}
\label{eq:sigmaT_rep}
\end{equation}
To our knowledge, no closed expressions of this form exist for $\sigma_\mathrm{V}$.

\section{Detailed calculations for an attractive potential and large $\beta$}

\subsection{Distance of closest approach}
\label{app:Rmin}

As mentioned in Sec.~\ref{sec:largebeta_att}, the equation
\begin{align}
\label{eq:ueff2}
U_\mathrm{eff}(R) \equiv  \left(\frac{\ell + 1/2}{\kappa}\right)^2\frac{1}{R^2} - 2\beta\frac{e^{-R}}{R} =  1
\end{align}
can have multiple roots, the largest of which determines the distance of closest approach, $R_0$. The existence of multiple roots implies the presence of a local maximum at $R > 0$, for which $U_\mathrm{eff}^\prime(R) = 0$ or
\begin{equation}
\label{eq:ueffprime}
e^{-R} = \frac{1}{\beta}\left(\frac{\ell + 1/2}{\kappa}\right)^2 \frac{1}{R(R+1)}\,.
\end{equation}
For a given $\beta$ we can find the smallest $\ell$ for which multiple roots exist, by simultaneously solving Eqs.~(\ref{eq:ueff2}) and~(\ref{eq:ueffprime}), which yields
\begin{equation}
\label{eq:lab1}
\left(\frac{\ell + 1/2}{\kappa}\right) = R_0\sqrt{\frac{R_0+1}{R_0-1}}\,.
\end{equation}
For $R_0 \gtrsim1$, one can expand the equation above to get
\begin{align}
\frac{\ell + 1/2}{\kappa} = 1 + R_0 + \frac{1}{2R_0} \; ,
\end{align}
which implies
\begin{align}
R_0 \approx \frac{1}{2}\left(-1 + \frac{\ell+1/2}{\kappa} + \sqrt{-1 -\frac{2\ell + 1}{\kappa} + \left(\frac{\ell+1/2}{\kappa}\right)^2}\right)\,.
\end{align}
Substituting this value of $R_0$ back in Eq.~(\ref{eq:ueffprime}), we get
\begin{equation}
\ell_\mathrm{max} = \kappa\,\left(1 +\log\beta - \frac{1}{2\,\log\beta}\right)-\frac{1}{2}\;.
\end{equation}

\subsection{Derivation of phase shifts}
\label{app:phase_shifts}

We start from the integral
\begin{align}
    I^\ell & = \kappa \int_{R_0(\ell)}^\infty 
    \mathrm{d}{R} \sqrt{1 + \frac{2 \beta}{R} e^{-R} - \frac{(\ell+\tfrac{1}{2})^2}{\kappa^2 R^2}} \\
    & = \kappa \int_{R_0(\ell)}^{R_\text{cut}} 
    \mathrm{d}{R} \sqrt{1 + \frac{2 \beta}{R} e^{-R} - \frac{(\ell+\tfrac{1}{2})^2}{\kappa^2 R^2}} + \kappa \int_{R_\text{cut}}^\infty
    \mathrm{d}{R} \sqrt{1 + \frac{2 \beta}{R} e^{-R} - \frac{(\ell+\tfrac{1}{2})^2}{\kappa^2 R^2}} \\
    & \equiv I_1^\ell + I_2^\ell\; ,
\end{align}
where $R_0 = (\ell + 1/2)^2 / (2 \beta \kappa^2)$ and $R_\text{cut}$ is chosen such that $R_0 \ll R_\text{cut} \ll 1$.

In the first integral we can neglect the first term in the square root and approximate $e^{-R} \approx 1$, leading to
\begin{align}
    I^\ell_1 & = \kappa \int_{(\ell+\tfrac{1}{2})^2 / (2 \beta \kappa^2)}^{R_\text{cut}} 
    \mathrm{d}{R} \sqrt{\frac{2 \beta}{R}- \frac{(\ell+\tfrac{1}{2})^2}{\kappa^2 R^2}} \\
    & = (\ell+\tfrac{1}{2}) \int_{1}^{S_\text{cut}} 
    \mathrm{d}{S} \frac{1}{S} \sqrt{S - 1} \\
    & = (2 \ell + 1) \left(\sqrt{S_\text{cut} -1} - \text{asec}(\sqrt{S_\text{cut}})\right)\;,
\end{align}
with $S_\text{cut} = R_\text{cut} (2 \beta \kappa^2) / (\ell+\tfrac{1}{2})^2$. For $S_\text{cut} \gg 1$ we then find
\begin{equation}
    I^\ell_1 - I^{\ell-1}_1 = -\pi + \sqrt{\frac{2}{\beta R_\text{cut}}}  \frac{\ell}{\kappa} \; .
\end{equation}

In the second integral, we have $(\ell+\tfrac{1}{2})^2 \ll \kappa^2 R^2$, such that we can approximate
\begin{equation}
I^\ell_2 \approx \kappa \int_{R_\text{cut}}^\infty
    \mathrm{d}{R} \sqrt{1 + \frac{2 \beta}{R} e^{-R}} - \frac{(\ell+1/2)^2}{2 \kappa^2} \frac{1}{\sqrt{1 + \tfrac{2 \beta}{R} e^{-R}} R^2} \; .
\end{equation}
We then find
\begin{equation}
    I^\ell_2 - I^{\ell-1}_2 \approx - \frac{\ell}{\kappa}\int_{R_\text{cut}}^\infty
    \mathrm{d}{R} \frac{1}{\sqrt{1 + 2 \beta e^{-R} / R} R^2}\;.
\end{equation}
To solve this integral, we again split it into two parts, separated at $R = R_\text{t}$ defined by
$2 \beta e^{-R_\text{t}} / R_\text{t} = 1$, which can equivalently be written as $R_\text{t} = W(2 \beta)$:
\begin{align}
    I^\ell_2 - I^{\ell-1}_2 & \approx - \frac{\ell}{\kappa}\int_{R_\text{cut}}^{R_\text{t}}
    \mathrm{d}{R} \frac{1}{\sqrt{1 + \tfrac{2 \beta}{R} e^{-R}} R^2} - \frac{\ell}{\kappa}\int_{R_\text{t}}^\infty
    \mathrm{d}{R} \frac{1}{\sqrt{1 + \tfrac{2 \beta}{R} e^{-R}} R^2} \\
    & \approx - \frac{\ell}{\kappa}\int_{R_\text{cut}}^{R_\text{t}}
    \mathrm{d}{R} \frac{1}{\sqrt{2 \beta e^{-R}} R^{3/2}} - \frac{\ell}{\kappa}\int_{R_\text{t}}^\infty
    \mathrm{d}{R} \frac{1}{R^2} \\
    & \approx - \frac{\ell}{\kappa} \left[\sqrt{\frac{2}{R_\text{cut} \beta}} - \sqrt{\frac{2 e^{R_\text{t}}}{R_\text{t} \beta}} + \sqrt{\frac{\pi}{\beta}} \text{erfi}(\sqrt{R_\text{t} / 2}) + \frac{1}{R_\text{t}} \right] \; ,    
\end{align}
where $\text{erfi}(x)$ denotes the imaginary error function. For $R_\mathrm{t} \gg 1$ we can approximate
\begin{equation}
 \sqrt{\pi}\text{erfi}\left(\sqrt{R_\mathrm{t}/2}\right) \approx \sqrt{\frac{2}{R_\mathrm{t}}} e^{R_\mathrm{t}/2} \left(1 + \frac{1}{R_\mathrm{t}}\right) \; .
\end{equation}
Now we make use of the defining property of $R_\mathrm{t}$ to substitute $e^{R_\mathrm{t}} = 2\beta/R_\mathrm{t}$ and obtain
\begin{equation}
    I^\ell_2 - I^{\ell-1}_2 \approx - \frac{\ell}{\kappa}\left[\sqrt{\frac{2}{R_\text{cut} \beta}} + \frac{R_\mathrm{t} + 2}{R_\mathrm{t}^2} \right] \; . 
\end{equation}
As expected, the dependence on $R_\text{cut}$ drops out when combining $I_1$ and $I_2$, giving
\begin{equation}
I^\ell - I^{\ell-1} = -\pi - \frac{\ell}{\kappa} \frac{W(2 \beta) + 2}{W(2 \beta)^2}  \equiv -\pi- \frac{\ell}{\kappa} \gamma(\beta)\;.
\end{equation}
Adding $\pi/2$ following Eq.~\eqref{eq:deltalexact} yields Eq.~\eqref{eq:ddeltal_att}.

\section{Velocity averaging}
\label{app:averaging}

In this appendix we discuss the role of the momentum transfer cross section and the viscosity cross section for describing the effects of DM self-interactions and derive the appropriate expressions for evaluating these cross sections for a given distribution of DM velocities. 

We first consider a DM particle moving with velocity $v$ through a background density $\rho$ of DM particles at rest. The particle encounters a flux of DM particles given by $\rho \, v / m_\chi$. The probability for scattering to occur is given by $\sigma \rho \, v / m_\chi$. In the center-of mass (cms) frame (which moves with velocity $v/2$) the distribution of scattering angles is given by $\sigma^{-1} \mathrm{d}\sigma/\mathrm{d}\theta_\text{cms}$. For a given scattering angle $\theta_\text{cms}$, the velocity transfer is given by $\Delta \mathbf{v} = \mathbf{v}_\text{cms} - \mathbf{v}'_\text{cms}$, where $\mathbf{v}_\text{cms}$ and $\mathbf{v}'_\text{cms}$ denote the velocity of the incoming DM particle in the cms frame before and after the collision, respectively. We decompose $\Delta \mathbf{v}$ into a component parallel and a component perpendicular to $\mathbf{v}$: $\Delta v_\parallel = v_\text{cms} (1 - \cos \theta_\text{cms})$, $\Delta \mathbf{v}_\perp = - v_\text{cms} (\sin \theta_\text{cms} \cos \phi, \sin \theta_\text{cms} \sin \phi)$, where $\phi$ is the azimuthal angle. The expectation value of $\Delta v_\parallel$ is then given by
\begin{equation}
 \langle \Delta v_\parallel \rangle = \frac{v_\text{cms}}{\sigma} \int \frac{\mathrm{d}\sigma}{\mathrm{d}\theta_\text{cms}} (1 - \cos \theta_\text{cms}) \mathrm{d}\theta_\text{cms} = \frac{\sigma_\mathrm{T} v}{2 \, \sigma} \; ,
\end{equation}
where in the last step we have made use of the fact that $v_\text{cms} = v/2$. Since $\phi$ is distributed uniformly, the expectation value of $\Delta \mathbf{v}_\perp$ vanishes, but we can calculate the expectation value of $\Delta v_\perp^2$:
\begin{equation}
 \langle \Delta v_\perp^2 \rangle = \frac{v_\text{cms}^2}{\sigma} \int \frac{\mathrm{d}\sigma}{\mathrm{d}\theta_\text{cms}} \sin^2 \theta_\text{cms} \mathrm{d}\theta_\text{cms} = \frac{\sigma_\mathrm{V} v^2}{4 \, \sigma} \; .
\end{equation}
We can now calculate the relative rate of change of the momentum $p$,
\begin{equation}
\frac{\dot{p}}{p} = \frac{\rho}{m_\chi} \frac{v \sigma_\mathrm{T}}{2}\;,
\end{equation}
and of the energy in the transverse direction $E_\perp$,
\begin{equation}
\frac{\dot{E_\perp}}{E} = \frac{\rho}{m_\chi}\frac{v \sigma_\mathrm{V}}{4} \; .
\end{equation}

Let us now consider the case that the background DM particles are not at rest but have a velocity distribution $g(\mathbf{w})$. We assume that $g(\mathbf{w})$ is isotropic and hence we can take $\mathbf{v}$ to point in the $z$ direction. We begin by considering only particles with a specific velocity $\mathbf{w}$. The rate at which the incoming particle encounters such particles is given by $\rho v_\text{rel} / m_\chi$, where $\mathbf{v}_\text{rel} = \mathbf{v} - \mathbf{w}$. Following the same argument as above (and noting that now $v_\text{cms} = v_\text{rel}/2$), scattering will on average lead to a velocity transfer of $\langle \Delta v_\parallel \rangle = \frac{\sigma_\mathrm{T}}{\sigma} \frac{v_\text{rel}}{2}$, pointing in the direction of $\mathbf{v}_\text{rel}$. The relative change in momentum is therefore given by
\begin{equation}
 \frac{\dot{p}}{p} = \frac{\rho}{m_\chi} \frac{v_\text{rel}^2 \sigma_\mathrm{T}(v_\text{rel})}{2 v} \; ,
\end{equation}
which differs from the result above by a factor of $v_\text{rel}/v$. This additional factor is a direct consequence of the fact that the change in velocity in a single collision cannot exceed $v_\text{rel}$. The smaller $v_\text{rel}$, the more collisions are necessary to change the velocity of the incoming particle by a relevant amount. For the case of the perpendicular energy, one obtains
\begin{equation}
 \frac{\dot{E_\perp}}{E} = \frac{\rho}{m_\chi} \frac{v_\text{rel}^3 \sigma_\mathrm{V}}{4 v^2} \; .
\end{equation}

We now apply these results to the case where both scattering DM particles are bound to the same DM halo. In this case both $v$ and $w$ approximately follow a Maxwell-Boltzmann distribution with velocity dispersion $v_0$:
\begin{equation}
 f(v) = \sqrt{\frac{2}{\pi}} \frac{v^2 e^{-v^2/(2 v_0^2)}}{v_0^3} \; .
\end{equation}
The typical momentum is then given by $\langle p \rangle = 2 \sqrt{2/\pi} m_\chi v_0 $ and the typical energy is $\langle E \rangle = \tfrac{3}{2} m_\chi v_0^2$. The relative velocity $v_\text{rel}$ follows a Maxwell-Boltzmann distribution with dispersion parameter $\sqrt{2} v_0$. The expected relative change in momentum is then
\begin{equation}
\Gamma_p \equiv  \frac{\langle \dot{p} \rangle}{\langle p \rangle} = \frac{\rho}{m_\chi} \frac{\langle \sigma_\mathrm{T} v_\text{rel}^2 \rangle}{4 \sqrt{2/\pi} v_0} = \frac{\rho}{m_\chi} \int\mathrm{d}v_\mathrm{rel} \sigma_\mathrm{T}(v_\text{rel}) \frac{v_\text{rel}^4}{8 \sqrt{2} v_0^4} e^{-v_\text{rel}^2/(4 v_0^2)} \; ,
\end{equation}
while the expected relative change in energy in the transverse direction is
\begin{equation}
\Gamma_E \equiv  \frac{\langle \dot{E_\perp} \rangle}{\langle E \rangle} = \frac{\rho}{m_\chi} \frac{\langle \sigma_\mathrm{V} v_\text{rel}^3 \rangle}{6 v_0^2} = \frac{\rho}{m_\chi} \int \mathrm{d}v_\mathrm{rel}\sigma_\mathrm{V}(v_\text{rel}) \frac{v_\text{rel}^5}{12 \sqrt{\pi} v_0^5} e^{-v_\text{rel}^2/(4 v_0^2)} \; .
\end{equation}
By defining
\begin{align}
\overline{\sigma_\mathrm{T}} = \frac{\langle \sigma_\mathrm{T} v_\text{rel}^2\rangle}{16 \sqrt{2} v_0^2 / \pi} \, , \qquad 
\overline{\sigma_\mathrm{V}} = \frac{\langle \sigma_\mathrm{V} v_\text{rel}^3 \rangle}{24 / \sqrt{\pi} v_0^3} \; ,
\end{align}
we can write these rates as
\begin{align}
\Gamma_p = \frac{\rho}{m_\chi}  \overline{\sigma_\mathrm{T}} \, \langle v_\text{rel} \rangle \, , \qquad
\Gamma_E = \frac{\rho}{m_\chi} \overline{\sigma_\mathrm{V}} \, \langle v_\text{rel} \rangle \; .
\end{align}

Finally, let us consider the case in which the cross sections can be written as
\begin{equation}
    \sigma_\mathrm{T,V} = \frac{\pi}{m_\phi^2} f_\mathrm{T,V}(\beta, \kappa) \; .
\end{equation}
The averaged cross sections are then given by
\begin{align}
\overline{\sigma_\mathrm{T}} & = \frac{\pi}{m_\phi^2} \int \mathrm{d} x
e^{-x^2/4} \frac{x^4}{32 \sqrt{2 / \pi}} f_\mathrm{T}(\beta_0 / x^2, \kappa_0 x)\, ,
 \nonumber \\
\overline{\sigma_\mathrm{V}} & = \frac{\pi}{m_\phi^2} \int \mathrm{d} x
e^{-x^2/4} \frac{x^5}{48} f_\mathrm{V}(\beta_0 / x^2, \kappa_0 x)
 \; ,
\end{align}
where $x = v / v_0$, and $\beta_0$ and $\kappa_0$ are the effective parameters obtained when setting $v = v_0$. For $\kappa_0 \gg 1$ the velocity averaged cross section hence depends only on $\beta_0$ and (trivially) on $m_\phi$. 

\bibliography{main}
\bibliographystyle{apsrev4-1}
\end{document}